\documentclass[]{aa}  

\usepackage{graphicx}
\usepackage{graphicx}	
\usepackage{amsmath}	
\usepackage{amssymb}	
\usepackage{color}
\usepackage{xcolor}
\usepackage{longtable}
\usepackage{soul}

\usepackage{natbib}
\bibpunct{(}{)}{;}{a}{}{,} 

\DeclareRobustCommand{\ferre}{FER\reflectbox{R}E}	
\DeclareRobustCommand{\gaia}{\textit{Gaia} }	
\DeclareRobustCommand{\kms}{\,km\,s$^{-1}$}

\begin{document}

   \title{OCCASO IV.}

   \subtitle{Radial Velocities and Open Cluster Kinematics\footnote{Table \ref{tab:indiv_stars} is only available in electronic form
at the CDS via anonymous ftp to cdsarc.u-strasbg.fr (130.79.128.5)
or via http://cdsweb.u-strasbg.fr/cgi-bin/qcat?J/A+A/}}

   \author{R. Carrera\inst{1}
   \and
L. Casamiquela\inst{2}
\and
J. Carbajo-Hijarrubia\inst{3}
\and
L. Balaguer-N\'u\~nez\inst{3}
\and
C. Jordi\inst{3}
\and
M. Romero-G\'omez\inst{3}
\and
S. Blanco-Cuaresma\inst{4}
\and
T. Cantat-Gaudin\inst{3,5}
\and
J. Lillo-Box\inst{6}
\and
E. Masana\inst{3}
\and
E. Pancino\inst{7,8}
}

   \institute{ INAF-Osservatorio Astronomico di Padova, vicolo Osservatorio 5, 35122, Padova, Italy\\
              \email{jimenez.carrera@inaf.it}
         \and
             Laboratoire d’Astrophysique de Bordeaux, Univ. Bordeaux, CNRS, B18N, allée Geoffroy Saint-Hilaire, 33615 Pessac, France\\
        \and 
        Institut de Ci\`encies del Cosmos, Universitat de Barcelona (IEEC-UB), Mart\'i i Franqu\`es 1, E-08028 Barcelona, Spain\\
        \and
        Harvard-Smithsonian Center for Astrophysics, 60 Garden Street, Cambridge, MA 02138, USA
        \and
        Max-Planck-Institut f\"ur Astronomie, K\"onigstuhl 17, D-69117, Heidelberg, Germany
        \and
        Centro de Astrobiolog\'{\i}a (CAB, CSIC-INTA), Depto. de Astrof\'{\i}sica, ESAC campus 28692 Villanueva de la Ca\~nada (Madrid),
        \and
        INAF – Osservatorio Astrofisico di Arcetri, Largo E. Fermi 5, I-50125 Florence, Italy\\
        \and
Space Science Data Center – ASI, Via del Politecnico SNC, I-00133 Roma, Italy\\
             }

   \date{Received September 15, 1996; accepted March 16, 1997}

 
  \abstract
   {Open clusters (OCs) are widely used as test particles to investigate a variety of astrophysical phenomena from stellar evolution to Galactic evolution. \gaia and the complementary massive spectroscopic surveys are providing an unprecedented wealth of information about these systems.}
   {The Open Cluster Chemical Abundances from Spanish Observatories (OCCASO) survey aims to complement all this work by determining OCs accurate radial velocities and chemical abundances from high-resolution, R$\geq$60\,000, spectra.}
   {Radial velocities have been obtained by cross-correlating the observed spectra with a library of synthetic spectra which covers from early M to A spectral types. }
   {We provide radial velocities for 336 stars including several \gaia Benchmark Stars and objects belonging to 51 open clusters. The internal uncertainties of the derived radial velocities go from 10\,m\,s$^{-1}$ to 21\,m\,s$^{-1}$ as a function of the instrumental configuration used. The derived radial velocities, together with the \gaia proper motions, have been used to investigate the cluster membership of the observed stars. After this careful membership analysis we obtain average velocities for 47 open clusters. To our knowledge this is the first radial velocity determination for 5 of these clusters.  Finally, the radial velocities, proper motions, distances and ages have been used to investigate the kinematics of the observed clusters and in the integration of their orbits.}
   {}


   \maketitle
%

\section{Introduction}

Open clusters (OCs) are groups from several hundreds to tens of thousands of gravitationally bound stars located in the Galactic disc. Unlike more massive and complex globular clusters, all the stars of a given OC seem to share the same properties such as age, kinematics and initial chemical composition \citep{friel2013}. Open clusters cover a wide range of masses, luminosities, structural characteristics and ages. All together has motivated their use as probes of a large variety of astrophysical phenomena. They have been key laboratories to understand the stellar interiors, nucleosynthesis and evolution in issues such as convection and radiation transport. Moreover, because of the wide ranges of ages covered and their distribution well around the disc they have been fundamental in the study of the Galactic disc evolution, both chemical and dynamical.  

The astrometric \gaia mission \citep{2016A&A...595A...1G} and the complementary ground-based spectroscopic surveys are causing a revolution in our knowledge of the Milky Way and its companions dwarf galaxies including, of course, open clusters. Until now, \gaia has sampled more than 1.8 billion stars \citep{gaia_edr3brown}. Its unprecedented accurate positions, $\alpha$ and $\delta$, proper motions, $\mu_{\alpha}$ and $\mu_{\delta}$, and parallaxes, $\varpi$, have allowed to significantly improve the membership determination causing the discovery of new systems \citep[e.g.][]{cantatgaudin2018,castroginard2018,castroginard2019,castroginard2020} but also to investigate their extension further away of their tidal radius \citep[e.g.][]{carrera2019m67}. Additionally, \gaia also measures magnitudes from three photometric bands \textit{G, $G_{BP}$} and \textit{$G_{RP}$} providing an unprecedented homogeneous photometric database \citep{gaia_edr3riello}.

Moreover, \gaia measures radial velocities \citep{brown_gaiadr2} for bright objects, 4$\leq$\textit{G}$\leq$13\,mag, and in the future it will determine chemical abundances for a handful of elements. This has motivated the development of several ongoing and forthcoming complementary massive ground-based spectroscopic surveys. These surveys are
GES-GIRAFFE \citep[Gaia-ESO Survey,][]{ges_gilmore,ges_randich}, APOGEE \citep[Apache Point Observatory Galactic Evolution Experiment,][]{apogee}, GALAH \citep[GALactic Archaeology with HERMES,][]{galah_dr2}, and the forthcoming WEAVE \citep[WHT Enhanced Area Velocity Explorer,][]{}, 4MOST \citep[4-metre Multi-Object Spectroscopic Telescope,][]{4most} and MOONS \citep[Multi Object Optical and Near-infrared Spectrograph for the VLT,][]{moons}  Galactic surveys. All together, they are going to measure radial velocities and chemical abundances for more than 1 million stars. However, most of them are sampling with an intermediate-resolution, R$\sim$20\,000, specific windows in the visible, e.g. GES-GIRAFFE, GALAH, WEAVE and 4MOST; or the \textit{H}-band in the infrared such as APOGEE or MOONS. In addition, GES-UVES is using the UVES \citep[Ultraviolet and Visual Echelle Spectrograph,][]{uves} instrument with an spectral resolution of R$\sim$47\,000 covering a wavelength range between 480 and 700\,nm.

These surveys need a complementary high-resolution, R$\geq$60\,000, spectroscopy with a larger wavelength coverage, e.g. 400 to 900\,nm allowing the determination of radial velocities and chemical abundances with a higher accuracy and precision that can be used as reference for the massive surveys described above. The large wavelength coverage allows measuring abundances for elements produced through all nucleosynthesis chains. This includes several elements which are not studied by the massive surveys, but provide robust constraints to the stellar evolutionary models and to the Galactic disc chemical history.

With the aim of providing this complementary high-resolution spectroscopic for OCs we are developing the OCCASO (Open Clusters Chemical Abundances from Spanish Observatories) survey \citep[see][for a detailed description, hereafter referred to as Paper\,I]{occaso1}. OCCASO was designed to investigate the Galactic disc chemical evolution \citep[see][hereafter referred to as Paper\,II, III, respectively]{occaso2,occaso3} from the point of view of OCs. For this reason, we mainly sample stars in the same evolutionary stage, the red-clump. These stars are among the brightest objects in these clusters, they can be easily identified even in the sparsely populated colour-magnitude diagrams, and their spectra are relatively less line-crowded and therefore, easier to analyse. The abundances determined from OCCASO spectra has allowed the robust detection of a genuine $\alpha$-enhancement in the young OC NGC~6705 with still not clear origin \citep{occaso_ngc6705}. The present paper is the fourth of the series directly based on OCCASO spectra, in which we present radial velocity determinations for all the stars observed until now.

This paper is organized as follows. The observational strategy and target selection is revised in Sect.~\ref{sec:occaso}. The data reduction is presented in Sect.~\ref{sec:data_reduction}. Sect.~\ref{sec:radial_velocity} shows the radial velocity determination together with internal and external comparisons. The average clusters radial velocities are derived in \ref{sec:avg_oc_vel}. The kinematic properties of studied OC are discussed on Sect.~\ref{sec:kinematic} and their orbits are obtained in Sect.~\ref{sec:orbits}. Finally, the main conclusions are presented in Sect.~\ref{sec:conclusion}

\begin{table}
\setlength{\tabcolsep}{0.9mm}
   \begin{center}
      \caption{Summary of instrumental configurations used for the OCCASO project.}
	\label{tab:instruments}
	\begin{tabular}{lcccc}
		\hline
		Code & Telescope & Instrument & Spectral & Wavelength\\
		 & & & Resolution & coverage [nm] \\
		\hline
		NOT1\tablefootmark{a} & NOT & FIES & 67\,000 & 400-725\\
		NOT2\tablefootmark{b} & NOT & FIES & 67\,000 & 400-900\\
		MERC & Mercator & HERMES & 85\,000 & 400-900\\
		CAH2\tablefootmark{c} & CAHA 2.2m & CAFE$_2$ & 62\,000 & 400-900\\
		\hline
	\end{tabular}
	\tablefoot{\tablefoottext{a}{We refer to the observations performed before 2017}\tablefoottext{b}{We refer to the observations performed from 2017}\tablefoottext{c}{We refer to the upgrade of the instrument performed in 2018 (see text for details).}}
 \end{center}
\end{table}

\section{Observational strategy and target selection}\label{sec:occaso}

\begin{table*}[h!]
\setlength{\tabcolsep}{1mm}
\begin{center}
\caption{Properties of the observed clusters.}
\begin{tabular}{lcccccccccccc}
\hline
  Cluster        & $\alpha_{\rm ICRS}$ & $\delta_{\rm ICRS}$ & $\varpi_{EDR3}$ & $\mu_{\alpha_{EDR3}}$ & $\mu_{\delta_{EDR3}}$ & Age & A$_V$ & distance & X & Y & Z & $R_{\rm GC}$\\
                 & [deg]    & [deg] & [mas] & [mas~a$^{-1})$] & [mas~a$^{-1}$]  & [Ga] & [mag] & [pc] & [pc] & [pc] & [pc] & [pc]\\
\hline
ASCC\,108 & 298.306 & 39.349         &  0.84$\pm$0.04 &  -0.47$\pm$0.09 &  -1.72$\pm$0.11  & 0.11 & 0.34 & 1160 & 310 & 1110 & 122 & 8105\\
Alessi\,1 & 13.343 & 49.536          &  1.42$\pm$0.03 &   6.49$\pm$0.09 &  -6.41$\pm$0.17  & 1.44 & 0.08 & 689 & -367 & 561 & -159 & 8726\\
Berkeley\,17 & 80.13 & 30.574        &  0.30$\pm$0.06 &   2.54$\pm$0.14 &  -0.36$\pm$0.13  & 7.24 & 1.63 & 3341 & -3325 & 252 & -214 & 11668\\
COIN-Gaia\,11 & 68.11 & 39.479       &  1.52$\pm$0.03 &   3.47$\pm$0.11 &  -5.70$\pm$0.12  & 0.79 & 1.25 & 669 & -635 & 199 & -68 & 8977\\
FSR\,0278 & 307.761 & 51.021         &  0.57$\pm$0.02 &  -4.58$\pm$0.12 &  -9.76$\pm$0.11  & 2.19 & 0.85 & 1708 & 63 & 1695 & 202 & 8448\\
FSR\,0850 & 86.257 & 24.74           &  0.45$\pm$0.05 &   1.27$\pm$0.12 &  -2.54$\pm$0.11  & 0.47 & 1.29 & 2232 & -2226 & -136 & -89 & 10567\\
IC\,4756 & 279.649 & 5.435           &  2.10$\pm$0.07 &   1.29$\pm$0.24 &  -4.98$\pm$0.26  & 1.29 & 0.29 & 506 & 406 & 299 & 47 & 7938\\
King\,1 & 5.505 & 64.383             &  0.54$\pm$0.05 &  -4.92$\pm$0.13 &  -1.32$\pm$0.13  & 3.63 & 1.9 & 1727 & -857 & 1499 & 51 & 9318\\
Melotte\,72 & 114.618 & -10.698      &  0.37$\pm$0.03 &  -4.14$\pm$0.07 &   3.69$\pm$0.07  & 0.98 & 0.41 & 2684 & -1793 & -1981 & 251 & 10325\\
Melotte\,111 & 186.014 & 25.652      & 11.65$\pm$0.37 & -12.11$\pm$0.59 &  -8.94$\pm$0.79  & 0.65 & 0 & 85 & -7 & -6 & 85 & 8347\\
NGC\,188 & 11.798 & 85.244           &  0.53$\pm$0.03 &  -2.32$\pm$0.13 &  -1.03$\pm$0.15  & 7.08 & 0.21 & 1698 & -851 & 1319 & 646 & 9285\\
NGC\,559 & 22.387 & 63.301           &  0.33$\pm$0.04 &  -4.28$\pm$0.10 &   0.17$\pm$0.12  & 0.26 & 2.19 & 2884 & -1743 & 2297 & 37 & 10341\\
NGC\,609 & 24.102 & 64.54            &  0.16$\pm$0.04 &  -0.69$\pm$0.07 &  -0.15$\pm$0.07  & 0.22 & 3.11 & 5023 & -3072 & 3969 & 183 & 12083\\
NGC\,752 & 29.223 & 37.794           &  2.26$\pm$0.05 &   9.73$\pm$0.25 & -11.82$\pm$0.27  & 1.17 & 0.07 & 483 & -324 & 303 & -191 & 8669\\
NGC\,1817 & 78.139 & 16.696          &  0.57$\pm$0.03 &   0.42$\pm$0.09 &  -0.94$\pm$0.08  & 1.12 & 0.59 & 1799 & -1742 & -189 & -405 & 10084\\
NGC\,1907 & 82.033 & 35.33           &  0.63$\pm$0.03 &  -0.15$\pm$0.11 &  -3.43$\pm$0.10  & 0.59 & 1.28 & 1618 & -1605 & 207 & 8 & 9947\\
NGC\,2099 & 88.074 & 32.545          &  0.67$\pm$0.04 &   1.87$\pm$0.18 &  -5.64$\pm$0.16  & 0.45 & 0.75 & 1432 & -1429 & 58 & 77 & 9769\\
NGC\,2126 & 90.658 & 49.883          &  0.76$\pm$0.03 &   0.80$\pm$0.13 &  -2.60$\pm$0.13  & 0.98 & 0.68 & 1304 & -1215 & 366 & 296 & 9563\\
NGC\,2266 & 100.832 & 26.976         &  0.26$\pm$0.04 &  -0.49$\pm$0.07 &  -1.27$\pm$0.10  & 0.81 & 0.26 & 3251 & -3169 & -433 & 581 & 11517\\
NGC\,2354 & 108.503 & -25.724        &  0.78$\pm$0.02 &  -2.86$\pm$0.10 &   1.86$\pm$0.12  & 1.41 & 0.35 & 1370 & -713 & -1158 & -163 & 9127\\
NGC\,2355 & 109.247 & 13.772         &  0.53$\pm$0.03 &  -3.83$\pm$0.11 &  -1.06$\pm$0.12  & 1.0 & 0.59 & 1941 & -1744 & -753 & 397 & 10112\\
NGC\,2420 & 114.602 & 21.575         &  0.39$\pm$0.03 &  -1.23$\pm$0.08 &  -2.03$\pm$0.09  & 1.74 & 0.04 & 2587 & -2316 & -757 & 869 & 10683\\
NGC\,2539 & 122.658 & -12.834        &  0.76$\pm$0.03 &  -2.32$\pm$0.11 &  -0.54$\pm$0.13  & 0.69 & 0.11 & 1228 & -713 & -971 & 236 & 9105\\
NGC\,2632 & 130.054 & 19.621         &  5.42$\pm$0.10 & -35.93$\pm$1.06 & -12.85$\pm$0.91  & 0.68 & 0 & 183 & -139 & -67 & 98 & 8479\\
NGC\,2682 & 132.846 & 11.814         &  1.16$\pm$0.04 & -11.00$\pm$0.19 &  -2.91$\pm$0.20  & 4.27 & 0.07 & 889 & -613 & -440 & 470 & 8964\\
NGC\,6603 & 274.616 & -18.409        &  0.32$\pm$0.04 &   0.16$\pm$0.10 &  -2.06$\pm$0.11  & 0.22 & 1.63 & 2727 & 2658 & 606 & -62 & 5713\\
NGC\,6633 & 276.845 & 6.615          &  2.53$\pm$0.05 &   1.28$\pm$0.31 &  -1.86$\pm$0.27  & 0.69 & 0.3 & 424 & 339 & 247 & 61 & 8004\\
NGC\,6645 & 278.158 & -16.918        &  0.55$\pm$0.04 &   1.34$\pm$0.10 &  -0.64$\pm$0.11  & 0.51 & 0.97 & 1810 & 1739 & 490 & -113 & 6618\\
NGC\,6705 & 282.766 & -6.272         &  0.41$\pm$0.05 &  -1.55$\pm$0.13 &  -4.17$\pm$0.13  & 0.31 & 1.2 & 2203 & 1955 & 1009 & -106 & 6464\\
NGC\,6728 & 284.715 & -8.953         &  0.53$\pm$0.04 &   1.32$\pm$0.09 &  -1.80$\pm$0.10  & 0.60 & 0.84 & 1829 & 1638 & 791 & -181 & 6747\\
NGC\,6755 & 286.942 & 4.224          &  0.40$\pm$0.04 &  -0.75$\pm$0.09 &  -3.56$\pm$0.10  & 0.14 & 2.34 & 2647 & 2069 & 1649 & -78 & 6483\\
NGC\,6791 & 290.221 & 37.778         &  0.21$\pm$0.05 &  -0.41$\pm$0.12 &  -2.27$\pm$0.11  & 6.31 & 0.7 & 4231 & 1423 & 3903 & 800 & 7942\\
NGC\,6811 & 294.34 & 46.378          &  0.87$\pm$0.03 &  -3.35$\pm$0.14 &  -8.82$\pm$0.11  & 1.07 & 0.09 & 1161 & 212 & 1116 & 241 & 8203\\
NGC\,6819 & 295.327 & 40.19          &  0.37$\pm$0.03 &  -2.88$\pm$0.11 &  -3.88$\pm$0.11  & 2.24 & 0.4 & 2765 & 754 & 2628 & 407 & 8027\\
NGC\,6939 & 307.917 & 60.653         &  0.53$\pm$0.02 &  -1.81$\pm$0.11 &  -5.47$\pm$0.11  & 1.70 & 0.85 & 1815 & -182 & 1764 & 386 & 8703\\
NGC\,6940 & 308.626 & 28.278         &  0.96$\pm$0.03 &  -1.94$\pm$0.15 &  -9.45$\pm$0.16  & 1.35 & 0.41 & 1101 & 376 & 1026 & -137 & 8029\\
NGC\,6991 & 313.621 & 47.4           &  1.77$\pm$0.05 &   5.65$\pm$0.36 &   8.42$\pm$0.30  & 1.55 & 0.2 & 577 & 26 & 576 & 15 & 8333\\
NGC\,6997 & 314.128 & 44.64          &  1.13$\pm$0.02 &  -4.26$\pm$0.14 &  -6.97$\pm$0.16  & 0.65 & 1.43 & 901 & 71 & 898 & -7 & 8317\\
NGC\,7142 & 326.29 & 65.782          &  0.41$\pm$0.03 &  -2.68$\pm$0.11 &  -1.36$\pm$0.09  & 3.09 & 1.16 & 2406 & -628 & 2288 & 396 & 9255\\
NGC\,7245 & 333.812 & 54.336         &  0.28$\pm$0.03 &  -3.94$\pm$0.09 &  -3.29$\pm$0.07  & 0.60 & 0.96 & 3210 & -632 & 3145 & -104 & 9507\\
NGC\,7762 & 357.472 & 68.035         &  1.03$\pm$0.02 &   1.46$\pm$0.17 &   3.97$\pm$0.16  & 2.04 & 1.91 & 897 & -408 & 794 & 91 & 8784\\
NGC\,7789 & -0.666 & 56.726          &  0.48$\pm$0.03 &  -0.91$\pm$0.12 &  -1.97$\pm$0.13  & 1.55 & 0.83 & 2100 & -901 & 1887 & -196 & 9432\\
Ruprecht\,171 & 278.012 & -16.062    &  0.63$\pm$0.04 &   7.69$\pm$0.14 &   1.08$\pm$0.14  & 2.75 & 0.68 & 1522 & 1458 & 430 & -82 & 6895\\
Skiff\,J1942+38.6 & 295.611 & 38.645 &  0.38$\pm$0.02 &  -3.12$\pm$0.05 &  -5.92$\pm$0.07  & 1.48 & 0.28 & 2378 & 700 & 2251 & 312 & 7964\\
UBC\,3 & 283.799 & 12.326            &  0.58$\pm$0.03 &  -0.57$\pm$0.05 &  -1.42$\pm$0.12  & 0.12 & 0.98 & 1704 & 1214 & 1187 & 141 & 7223\\
UBC\,6 & 344.01 & 51.187             &  0.69$\pm$0.02 &  -4.61$\pm$0.05 &  -4.98$\pm$0.11  & 0.74 & 0.64 & 1493 & -387 & 1428 & -199 & 8843\\
UBC\,44 & 31.11 & 54.359             &  0.37$\pm$0.03 &  -2.29$\pm$0.15 &  -0.27$\pm$0.17  & 0.81 & 0.51 & 2737 & -1871 & 1969 & -333 & 10399\\
UBC\,59 & 82.239 & 48.043            &  0.43$\pm$0.04 &   0.63$\pm$0.18 &  -2.00$\pm$0.17  & 0.49 & 0.95 & 2585 & -2439 & 789 & 334 & 10808\\
UBC\,106 & 280.475 & -5.417          &  0.40$\pm$0.03 &  -1.05$\pm$0.08 &  -1.36$\pm$0.10  & 0.16 & 1.92 & 2353 & 2096 & 1069 & -14 & 6334\\
UBC\,215 & 100.461 & -5.243          &  0.70$\pm$0.02 &  -0.13$\pm$0.07 &  -3.13$\pm$0.08  & 0.41 & 1.11 & 1419 & -1137 & -842 & -111 & 9514\\
UPK\,55 & 296.812 & 10.428           &  1.24$\pm$0.04 &  -1.43$\pm$0.10 &  -6.90$\pm$0.09  & 0.21 & 0.54 & 769 & 504 & 573 & -98 & 7856\\
\hline
\end{tabular}
	\tablefoot{The mean proper motions and parallaxes have been calculated using \gaia EDR3 assuming the cluster membership probabilities compiled by \citet{cantatgaudin2020} from \gaia DR2 (see text for details about the original source of each cluster). }
\label{tab:ocs}
\end{center}
\end{table*}

OCCASO is using three of the high-resolution spectroscopic facilities available at the Spanish observatories (see Table~\ref{tab:instruments}). These instrumental configurations allow us to obtain high-resolution (R$\geq$60\,000), and large wavelength coverage spectra, from optical, 400\,nm, to near-infrared, 900\,nm. The obtained spectra have a signal-to-noise ratio (S/N) above 50\,pix$^{-1}$ but typically above 70\,pix$^{-1}$. What that aim, we take three exposures per star with a minimum S/N of around 30\,pix$^{-1}$. FIES \citep[FIbre-fed Echelle Spectrograph,][]{fies} installed at the 2.5\,m NOT (Nordic Optical Telescope, La Palma, Spain) was upgraded in 2017 with a new fibre bundle and CCD detector which increase the wavelength coverage of the output spectra. For this reason, we handle the spectra acquired before and after the upgrade as spectra from two different instruments denoted as NOT1, before the upgrade, and NOT2, after the upgrade. Something similar happened with CAFE \citep[Calar Alto Fiber-fed Echelle spectrograph,][]{cafe} at the 2.2\,m CAHA (Centro Astron\'omico Hispano-Alem\'an, Almer\'{\i}a, Spain)  telescope, which was upgraded in 2018. In this case, the degradation of the instrument performance before the upgrade prevent us of using the spectra observed until that moment\footnote{Radial velocities determined from those spectra were reported in Paper\,I.}. The third instrument used is HERMES \citep[High-Efficiency and high-Resolution Mercator Echelle Spectrograph,][]{hermes} installed at the 1.2\,m Mercator Telescope (La Palma, Spain). 

More than 130 observing nights have been performed in the framework of the OCCASO survey and the observations will continue in the future. The initial target selection criteria is explained in detail in Paper\,I. Briefly, Northern Hemisphere OCs older than 0.3\,Ga were selected trying to homogeneously sample the ranges of ages, metallicities, distances above/below the plane and Galactocentric distances. Moreover, these systems must have six or more stars in the expected position of the red clump in order to have a reasonable statistics for each cluster.  Owing to the limitations of the telescopes/instruments used in OCCASO, our sample is constrained to clusters which red clump position is brighter than $V\sim$15\,mag. When OCCASO started in 2013, we used all the information available in the literature, e.g. colour-magnitude diagrams, radial velocities or proper motions, in order to select the most likely members for each cluster. This initial strategy has been revised once the \gaia second data release \citep[DR2,][]{brown_gaiadr2} has been available which has contributed to improve significantly the membership determination. First, we based our membership selection on the membership probabilities determined from \gaia DR2 proper motions and parallaxes by \citet{cantatgaudin2018}. Moreover, we included new systems discovered from \gaia DR2 with at least four stars at the expected position of the red clump: COIN-Gaia\, 11 \citep{cantat_gaudin2019coin}; UBC\,3 and UBC\,6 \citep{castroginard2018}; UBC\,44 and UBC\,59 \citep{castroginard2019}; and UBC\,106 and UBC\,215 \citep{castroginard2020}. In some cases where the red clump is so sparse, we also observed a few stars in the main-sequence in order to better constrain the cluster average radial velocity. Additionally, to the cluster targets we have also observed several of the \gaia FGK Benchmark Stars \citep[GBS,][]{heiter_gbs2015,gbs_jofre,gbs_blanco_cuaresma} in order to check our analysis methodology. Finally, we have performed observations of several stars with the different instrumental configurations in order to provide internal comparison.

\begin{figure*}
   \centering
   \includegraphics[width=\textwidth]{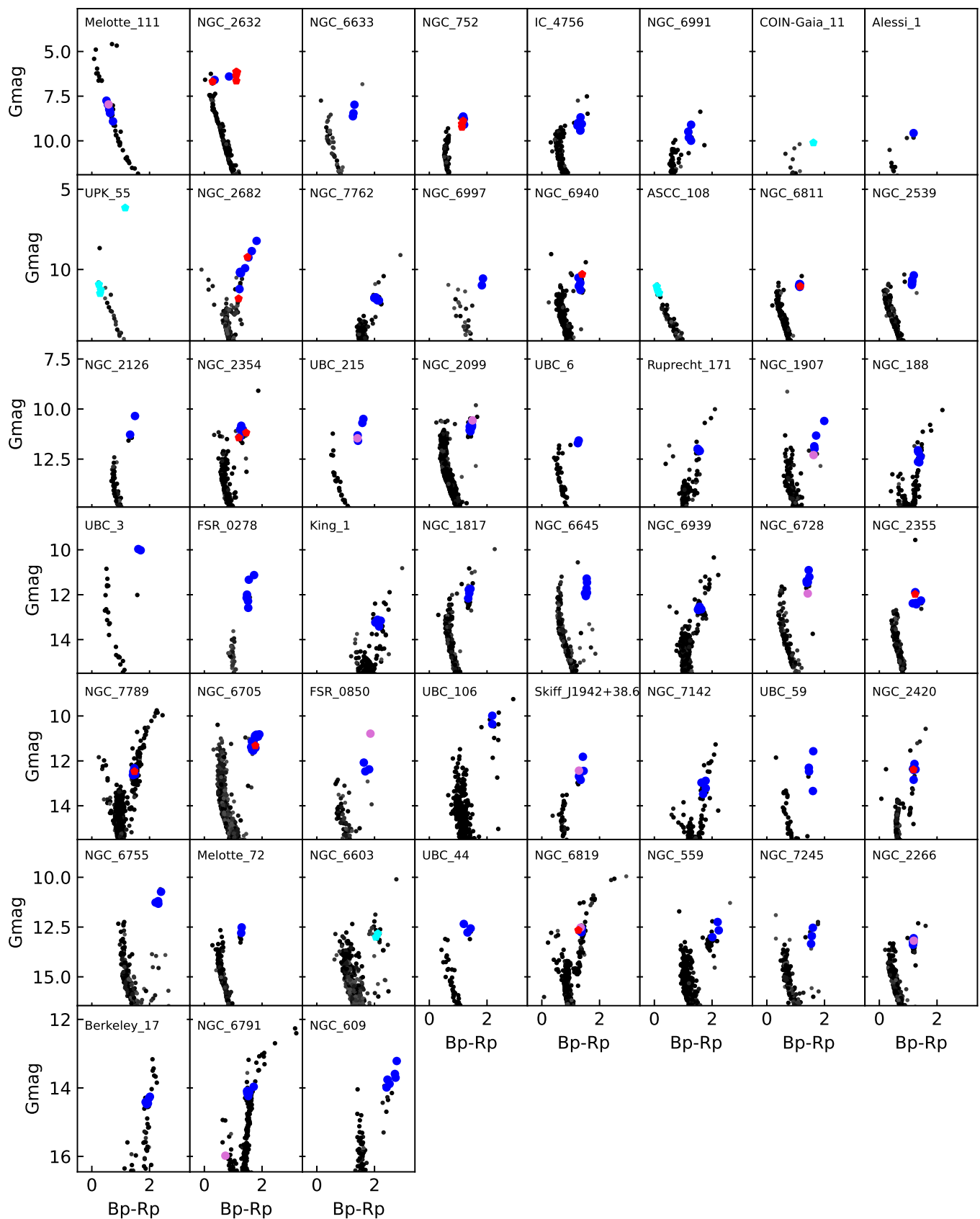}
   \caption{\gaia DR2 colour-magnitude diagrams of the observed clusters using member stars from \citet{cantatgaudin2020} (black points). Blue, orchid, and cyan points are the OCCASO spectroscopic targets considered as members, non-members, doubtful members, respectively (see text for details). Red points are spectroscopic binaries. Note that the clusters are sorted by distance from the Sun to improve understanding of the figure.}
    \label{fig:dcms}%
    \end{figure*}

In summary, at the moment OCCASO has sampled 312 stars in a total of 51 clusters. The sampled clusters are listed in Table~\ref{tab:ocs}. The mean proper motions and parallaxes have been obtained from the individual \gaia early third data release \citep[EDR3,][]{gaia_edr3brown} values, but using the cluster membership probabilities determined by \citet{cantatgaudin2020} based on \gaia DR2. The observed stars are listed in Table~\ref{tab:indiv_stars} and their positions in the colour-magnitude diagram of each cluster are shown in Fig.~\ref{fig:dcms}.

\section{Data reduction}\label{sec:data_reduction}

The data reduction strategy presented in Papers\,I and II has been fully revisited in order to make it fully automatic and to implement the radial velocity determination by cross-correlation. As explained in Paper\,I the bias subtraction, flatfield-correction, order tracing and extraction, and wavelength calibration is performed by dedicated pipelines specifically developed for each instrument: {\sl HERMESDRS} for HERMES \citep{hermes}; {\sl FIEStool} for FIES \citep{fies}; and {\sl CAFExtractor} for CAFE \citep{cafe_lillo}.

These pipelines also provide final spectra with all the orders merged. These 1D spectra were used in Paper\,I. Some wiggles were detected in these spectra, particularly in the overlapping regions between orders. These features do not affect the radial velocity determination, but they have a strong  impact on the abundance analysis because the continuum shape is distorted, and it is difficult to correct a posteriori with the normalization algorithm (see Paper\,II for details). For this reason, we now begin our data reduction procedure from the extracted  and  wavelength  calibrated  spectra,  but  still  separated  by  orders. From  here,  our  analysis  is  performed  by three modules. We describe the first two in Appendix~\ref{apex:data_reduction}, and the third one, which is devoted to determine radial velocities in Sect~\ref{sec:radial_velocity}. The entire code has been written in {\sl IDL} software language (Exelis Visual Information Solutions, Boulder, Colorado).

\section{Radial velocity determination}\label{sec:radial_velocity}

The radial velocities from the 1D averaged and order merged spectra have been obtained by measuring the Doppler velocity shifts of the spectral lines using the classical cross-correlation method \citep[e.g][]{tonrydavis}. To do that, the observed spectrum is cross-correlated against a template spectrum. The templates have been obtained from three coarse grids {\sl hnsc1}, {\sl hnsc2}, and {\sl hnsc3} described by \citet{allendegrids}. In global, we covered from early M, $T_{\rm eff}$=3500\,K, to A, $T_{\rm eff}$=12\,000\,K, spectral types, although most of our targets are GK-type. All these grids have three dimensions:  metallicity, [M/H]; effective temperature, $T_{\rm eff}$; and surface gravity, $\log g$. We refer the reader to \citet{allendegrids} for details about the ranges of the parameters covered by each grid. These grids have a resolution of 100\,000, although they were originally computed with an infinity resolution and 0.45\kms~sampling equivalent to a resolution of 300\,000. These grids have been smoothed to match the nominal resolution of each instrument listed in Table~\ref{tab:instruments}. 

The procedure followed to determine the radial velocity of each target is the following. (i) We performed an initial cross-correlation with a reference synthetic spectrum to obtain an initial shift for every star. In our case we used a spectrum with [M/H]=0.0\,dex, $T_{\rm eff}$ = 4500\,K, and $\log g$ = 2.0\,dex. (ii) After applying this initial shift, each averaged and order merged spectrum is compared with all the grids in order to identify the model parameters that best reproduces it. This step is performed with \ferre\footnote{Available at \url{https://github.com/callendeprieto}} \citep{allende_ferre}. \ferre~ selects the synthetic spectrum that better match each target from a $\chi^2$ minimization. (iii) The best-fitting synthetic spectrum is cross-correlated again with the observed spectrum in order to refine the shift between both. Steps (ii) and (iii) are repeated twice in order to refine the radial velocity determination. The derived radial velocities for each observed star and instrumental configuration are listed in Table~\ref{tab:indiv_stars}. This table includes a few objects with less than three individual exposures, one of our initial requirements. In most of the cases, this simply implies that the observations of these objects have not been concluded. We provide radial velocities because in most of the cases these are the first radial velocity determination from high-resolution spectra for these objects, although they are not used in our analysis.

\subsection{Radial velocity uncertainties}

\begin{figure}
	\includegraphics[width=\columnwidth]{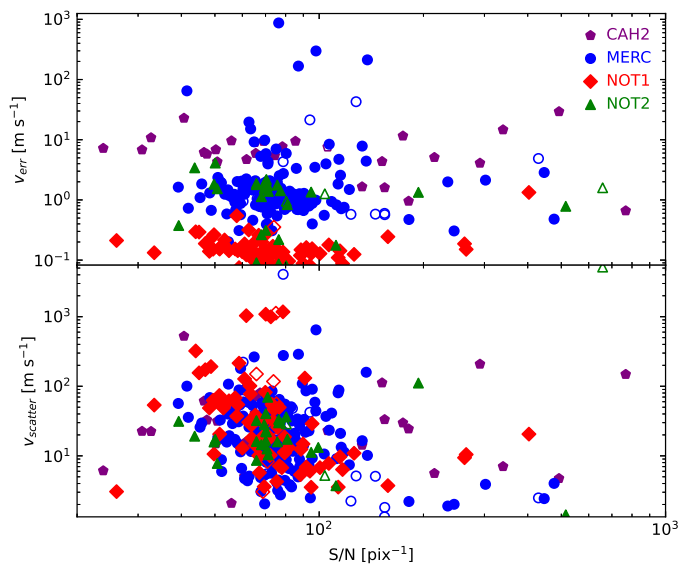}
    \caption{Variation of $v_{\rm err}$ (top) and $v_{\rm scatter}$ (bottom) as a function of S/N for the different telescopes/instruments used in our analysis. Closed and open symbols are single and spectroscopic binaries, respectively.}
    \label{fig:verr}
\end{figure}

Traditionally, the uncertainties of the radial velocities have been determined from the height of the cross-correlation peak \citep[see][for details]{tonrydavis}, here named as $v_{\rm err}$. This depends mainly on how well the template reproduces the averaged and order merged spectrum.  Top panel of Fig.~\ref{fig:verr} shows the run of $v_{\rm err}$ as a function of S/N for each telescope. There is no clear correlation with S/N, but it seems to be a dependence with the instrumental configuration. The lowest $v_{\rm err}$ values are obtained for NOT1 which has the lowest wavelength range coverage, 500-750\,nm. MERC and NOT2, which cover almost the same wavelength range, have a similar behaviour. CAH2, with a similar wavelength range than the previous two, has slightly larger $v_{\rm err}$. There is a group of objects observed with MERC which have large $v_{\rm err}$ values. This group is composed by early A-type stars in our sample, which typically have larger rotational velocities than the bulk of our sample formed by GK-type stars.

The $v_{\rm err}$ tends to underestimate the real uncertainties involved in the radial velocity determination. Owing to the observational strategy of OCCASO (see Sect.~\ref{sec:occaso}) at least three individual exposures are acquired for each object. Therefore, the radial velocity uncertainty can be determined in a more realistic way through the radial velocity scatter of the individual measurements for each star: $v_{\rm scatter}$. To derive $v_{\rm scatter}$ each order of each individual exposure is cross-correlated with the averaged one for every star. The shift between each individual exposure and the averaged one is obtained as the median of the shift found for each order. For a given exposure the shift found for each order does not show a significant dispersion. Bottom panel of Fig.~\ref{fig:verr} shows the run of $v_{\rm scatter}$ as a function of S/N for each telescope. Unlike $v_{\rm err}$, $v_{\rm scatter}$ does show a clear correlation with S/N and no relation to the instrumental configuration. Although our observations are not designed to detect spectroscopic binaries (because typically all the exposures of a given star are acquired one after the other) a few well known spectroscopic binaries, open symbols, tend to have larger $v_{\rm scatter}$ values than single stars. Therefore, large $v_{\rm scatter}$ may be related with binarity. The group of stars with $v_{\rm scatter}\sim$1\kms~observed with NOT1 (red symbols) are related to problems in one of our runs, Apr13. This problem was already reported in Paper\,I and is due to a poor wavelength calibration, which may be related to the use of inappropriate
calibration images when running the pipeline. The reduction with FIEStool of this run could
not be performed at the telescope, and it was
run a posteriori using a version built to be used outside the NOT facilities. We have tried to mitigate this by improving the wavelength calibration of this run. As consequence the initial offset of $\sim$5\kms~reported in Paper\,I has been reduced to $\sim$1\kms.

\begin{figure}
	\includegraphics[width=\columnwidth]{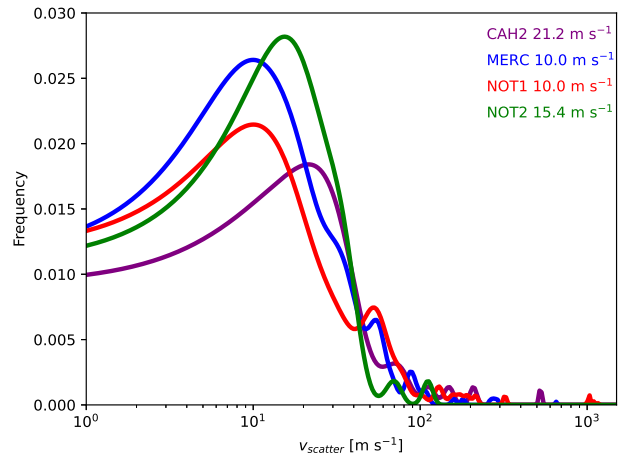}
    \caption{Histograms of the radial velocity scatter distribution for stars with three or more visits for the different telescopes/instruments used. The distributions peak at 21.2\,m\,s$^{\rm -1}$ for CAH2 (purple), 10.0\,m\,s$^{\rm -1}$ for MERC (blue), 10.3\,m\,s$^{\rm -1}$ for NOT1 (red), and 15.4\,m\,s$^{\rm -1}$ for NOT2 (green), respectively.}
    \label{fig:rvsactter_distribution}
\end{figure}

Figure~\ref{fig:rvsactter_distribution} shows the distributions of the $v_{\rm scatter}$ for each instrumental configuration. The distributions peak at 21.2\,m\,s$^{\rm -1}$ for CAH2, 10.0\,m\,s$^{\rm -1}$ for MERC, 10.3\,m\,s$^{\rm -1}$ for NOT1, and 15.4\,m\,s$^{\rm -1}$ for NOT2, respectively. This implies a significant improvement, about a factor 50, in the radial velocities uncertainties determination with respect to Papers\,I and II. 

\subsection{Internal comparison}

\subsubsection{Comparison between telescopes}

\begin{figure}
	\includegraphics[width=\columnwidth]{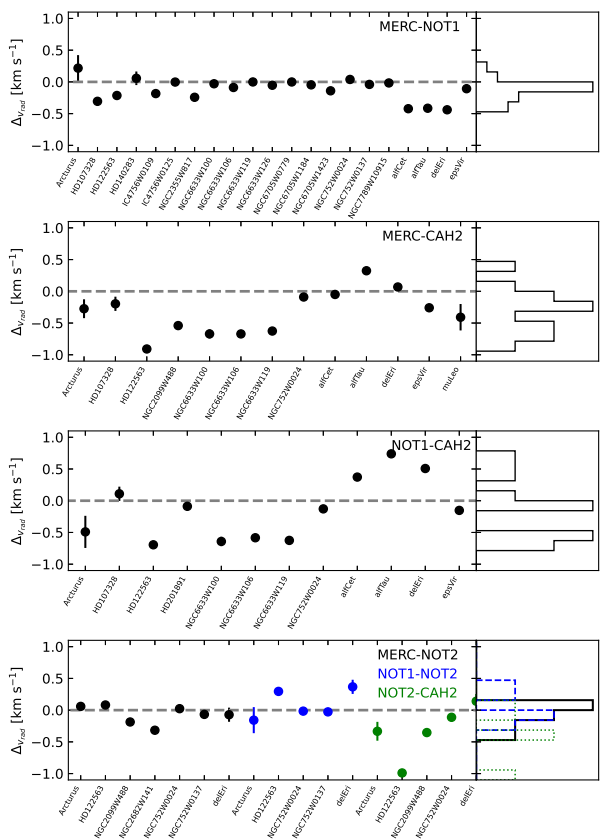}
    \caption{Differences in $v_{\rm rad}$ obtained for the stars in common between MERC and NOT1 (top panel), MERC and CAH2 (top central panel), NOT1 and CAH2 (bottom central panel), NOT2 and MERC (blue) and CAH2 (red), respectively (bottom panel). The error bars are the sum of the square of the uncertainties. In most of the cases, error bars are smaller than symbol sizes.}
    \label{fig:comp_telescopes}
\end{figure}

Several stars have been observed with the different instrumental configurations to evaluate the internal systematic errors. The comparison between the values obtained in different instrumental configurations are shown in Fig.~\ref{fig:comp_telescopes} and their statistics are summarized in Table~\ref{tab:vr_comp_config}. In general, there is good agreement, within the uncertainties, between the radial velocities derived from spectra acquired with MERC, NOT1 and NOT2 (in spite of the small number of stars observed in common between NOT2 and MERC, 6, and NOT1, 5). For this reason we consider that the radial velocities derived from the three instrumental configurations are on the same scale. 

In the case of CAH2, the derived radial velocities show significant differences when they are compared with the values obtained from the other instrumental configurations. We have no explanation for these differences. \citet{cafe_lillo} reported a dependence of the derived radial velocity with the S/N of the individual exposures. We have tried to take into account this effect using the relation provided by \citet{cafe_lillo} but this does not reduce significantly the differences. There is no clear systematic between the radial velocities derived from CAH2 and from the other telescopes as it is shown in Fig.~\ref{fig:comp_telescopes}, so we do not try to put all of them in the same scale. Therefore, we are going to use the CAH2 radial velocities in our analysis with care.

\begin{table}
\setlength{\tabcolsep}{1.25mm}
\begin{center}
	\caption{Differences of radial velocities measured with different instrumental configurations.}
	\label{tab:vr_comp_config}
	\begin{tabular}{lcccc} 
		\hline
		Instrumental & Median & SD & MAD & N\\
	    configuration & [\kms] & [\kms] & [\kms] &\\
		\hline
		MERC-NOT1 & -0.05 & 0.17 & 0.09 &  21\\
		MERC-NOT2 & -0.02 & 0.09 & 0.06 & 6\\
		MERC-CAH2 & -0.27 & 0.33 & 0.27 & 13\\
		NOT1-NOT2 & -0.01 & 0.20 & 0.14 & 5\\
		NOT1-CAH2 & -0.14 &  0.47 & 0.46 & 13\\
		NOT2-CAH2 & -0.33 & 0.37 & 0.22 & 5\\
	    \hline
	\end{tabular}
\end{center}
\end{table}

\subsubsection{Comparison with previous OCCASO radial velocities}

\begin{figure}
	\includegraphics[width=\columnwidth]{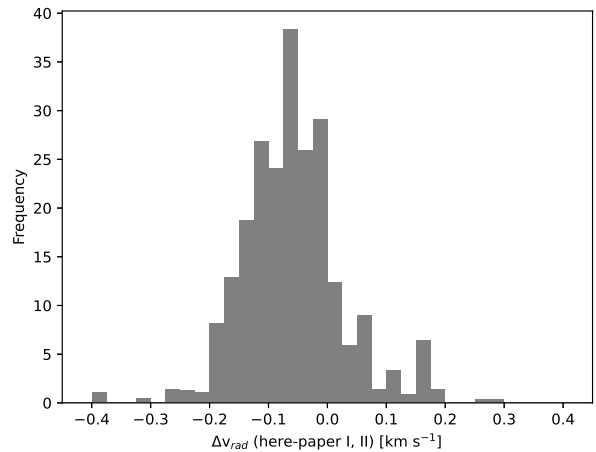}
    \caption{Distribution of the differences between $v_{\rm rad}$ derived in here and in Papers\,I and II.}
    \label{fig:rvcomp_oldpapers}
\end{figure}

Radial velocities in Papers\,I and II were derived using {\sl DAOSPEC} \citep{daospec} which is designed to determine equivalent widths of spectral lines. {\sl DAOSPEC} provides also a rough determination of the radial velocity by cross-matching the line centres with their reference rest wavelengths (see Paper\,I for details). Moreover, the final averaged and order merged spectra used in those publications were obtained for a slightly different procedure (see Sect.~\ref{sec:data_reduction}). In spite of the larger uncertainties involved in the radial velocities derived in Papers\,I and II, there is a very good agreement between both determinations for the 147 stars in common as shown in Fig.~\ref{fig:rvcomp_oldpapers} with a median of the differences of -0.05\kms~with a standard deviation of 0.09\kms~and a median absolute deviation of 0.05\kms.

There is only one star with a significant difference between both radial velocities determinations, the star \gaia EDR3 2194819856960726912. This star was flagged in Paper\,II as a non-member of NGC\,6939 since its radial velocity, $v_{\rm rad}$=-29$\pm$2\kms, was significantly different from the values derived for other stars in this cluster. However, the value obtained here from cross-correlation is $v_{\rm rad}$=-18.09$\pm$0.06\kms, in good agreement with the other NGC\,6939 member stars. As we used the same observations but a different reduction procedure, the observed discrepancy are due to problems in the previous reduction pipeline that were not detected. Moreover, the observed spectrum has a relatively low S/N of 39.2\,pix$^{\rm -1}$ in comparison with the other stars observed in this cluster. The cross-correlation is less sensitive to low S/N spectra than the cross-match of line centres performed by {\sl DAOSPEC}.

\subsection{Comparison with external catalogues}

In Paper\,I we performed a comparison with different literature sources. In most of the cases, these sources were focused on a single cluster. The differences changed significantly depending on the literature source. We refer the reader to Paper\,I for details. In the last years, thanks to the \gaia mission and the ground based spectroscopic surveys, there are a wealth of large samples with radial velocities determined homogeneously. These samples include OC stars in common with OCCASO. In this section we perform a comparison with all these surveys. We include also in this comparison the radial velocities determined in the framework of the WIYN Open Cluster Study \citep[WOCS, e.g.][]{wocs_geller2015_ngc2682} which has systematically sampled radial velocities for stars in several OCs in common with OCCASO. We also compare our radial velocities with the studies of \citet{nordstrom2004} and \citet{soubiran_rvstd}. The first one is the Geneva-Copenhagen survey which has measured radial velocities for about 13\,500 FG-type stars in the Solar neighbourhood. The former is the catalogue of radial velocity standard stars used to establish the radial velocity zero point  in \gaia DR2. In all these comparisons we exclude the radial velocities determined for CAH2 which, as discussed above, are more uncertain. Moreover, we use only stars with three or more individual exposures in our sample, and we exclude the previously known spectroscopic binaries. The differences of radial velocities with others surveys is in Table~\ref{tab:ext_comp} and shown in Fig.~\ref{fig:comp_surveys1}. A detailed discussion about the comparison with each sample can be found in Appendix~\ref{apex:comp_svry}. In brief, there is a good agreement between OCCASO and \gaia DR2 radial velocities in spite of the larger uncertainties involved in the \gaia measurements. For APOGEE DR16, WOCS, \gaia RVS standards \citep{soubiran_rvstd}, and \citet{mermilliod2008_RVOC} there is a small systematics between each of these samples and OCCASO but smaller than the involved uncertainties. For the remaining samples, the agreement is not particularly significant. Noticeable is the bimodal distribution found in the case of GES DR4 without a clear explanation (panel h of Fig~\ref{fig:comp_surveys1}). The distribution of the differences with LAMOST DR5, RAVE DR6, and in less degree for GALAH DR3 are unusual without a clear peak (panels e, i, and g of Fig.~\ref{fig:comp_surveys1}, respectively). The average radial velocities uncertainties involved in each sample range from the $\sim$5\kms~in the case of LAMOST to the $\sim$0.1\kms~for GALAH. They could explain the differences found in the case of LAMOST but certainly not in the case of GALAH.Unfortunately, the small number of objects in common between OCCASO and these samples, the best case is the 32 stars in common with LAMOST, prevent us to draw further conclusions.

\begin{table}
\setlength{\tabcolsep}{1.25mm}
\begin{center}
	\caption{Differences of radial velocities in the sense OCCASO minus other surveys, where N is the number of objects in common.}
	\label{tab:ext_comp}
	\begin{tabular}{lcccc} 
		\hline
		Survey & N & Median & MAD & SD\\
		 & & [\kms] & [\kms] & [\kms]\\
	\hline
        \gaia DR2 & 238 & -0.04 & 0.34 & 0.56\\
		\citet{mermilliod2008_RVOC} & 78 & 0.29 & 0.12 & 0.22\\
		APOGEE DR16 & 50 & -0.23 & 0.13 & 0.15 \\
		WOCS & 33 & 0.19 & 0.15 & 0.22\\
		LAMOST DR6 & 32 & 5.5 & 2.5 & 3.8\\
        \citet{soubiran_rvstd} & 28 & 0.22 & 0.03 & 0.06\\
		GALAH DR3 & 25 & -0.23 & 0.80 & 1.40\\
		GES DR4\tablefootmark{a} & 20 & 0.36 & 0.37 & 0.47 \\
		RAVE DR6 & 16 & -0.22 & 0.85 & 1.00\\
		\citet{nordstrom2004} & 13 & 0.64 & 0.30 & 0.45\\
		SEGUE DR12 & 5 & 0.55 & 0.22 & 0.43\\
		\hline
	\end{tabular}
		\tablefoot{\tablefoottext{a}We provide the values for the whole sample for the sake of homogeneity with the other surveys. The medians of the two peaks are -0.37 and 0.53\kms~with standard deviations of 0.19 and 0.15\kms, respectively (see panel h of Fig.~\ref{fig:comp_surveys1} and text for details.)}
\end{center}
\end{table}

\begin{figure}
	\includegraphics[width=\columnwidth]{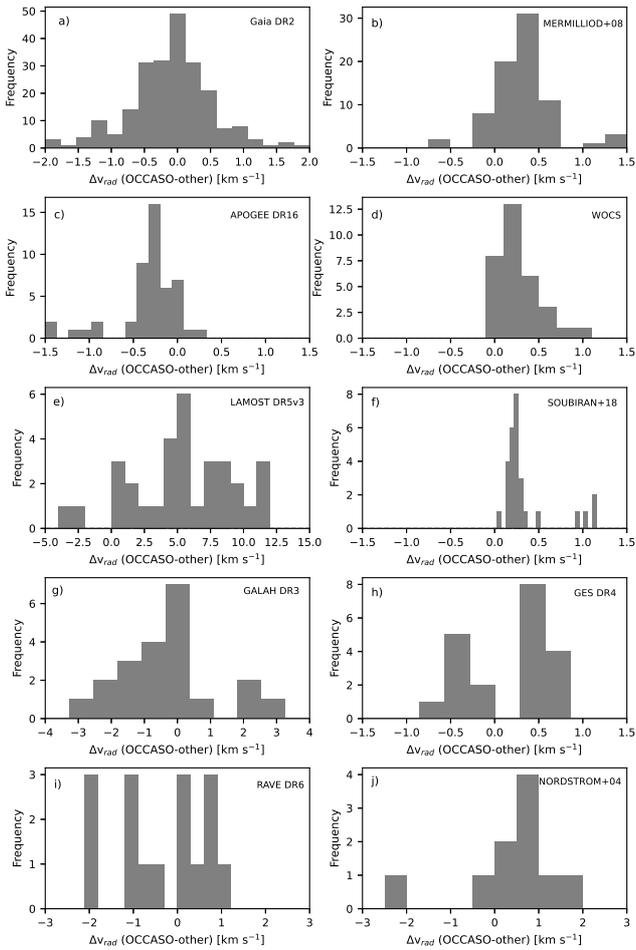}
    \caption{Comparison of the OCCASO radial velocities with different samples available in the literature.}
    \label{fig:comp_surveys1}
\end{figure}

\section{Open Clusters average radial velocities: membership}\label{sec:avg_oc_vel}

\begin{figure}[!]
\centering
\includegraphics[width=\columnwidth]{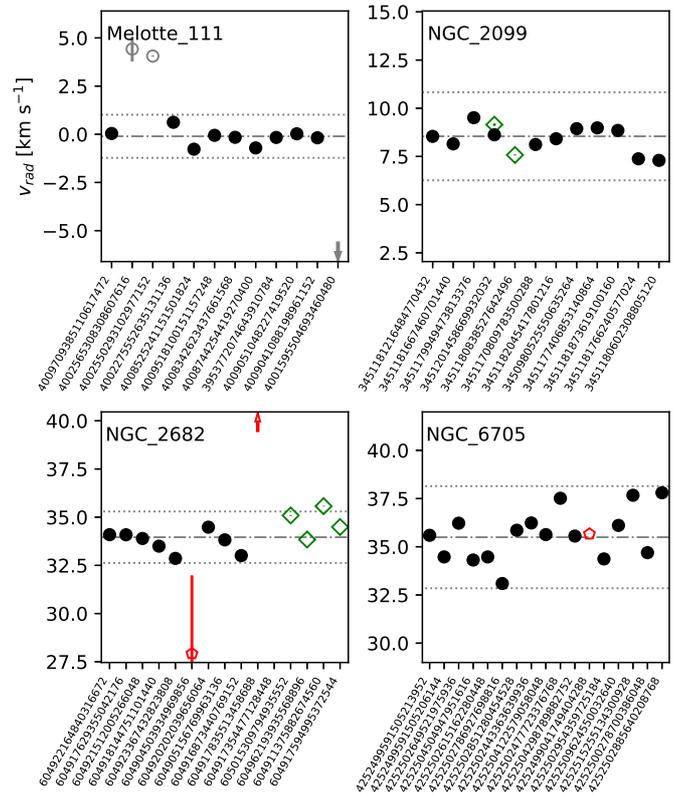}
\caption{Determination of the average radial velocity for each cluster. Filled symbols mark those stars used
to determine the average value, while open symbols are the excluded objects. Black circles represent those stars observed with MERC, NOT1 and NOT2, while green diamonds are the stars observed with CAH2. Red pentagons are the known spectroscopic binaries. Arrows denote objects outside the panels. Dot-dashed lines correspond to the average radial velocity for each cluster. Dotted lines show the $3\times\sigma_{v_{rad}}$ level. Note that in  most of the cases, error bars are smaller than the symbol size. All source identifications are from \gaia EDR3.}
\label{fig:rv_membership}%
\end{figure}

In order to determine the average radial velocity for each cluster, we have followed the same procedure described by \citet{soubiran2018gaiadr2_rv}. The average radial velocity is obtained using: 

\begin{equation}\label{eq:meanrv}
v_{rad,OC} = \frac{\sum_i v_{rad,i} \times w_i}{\sum_i w_i}
,\end{equation}
where $v_{rad,i}$ is the individual radial velocity for each star in the cluster and the weight $w_i$ is defined as $w_i=1/(v_{scatter,i})^2$. 

In the same way, the internal velocity dispersion is derived as:

\begin{equation}\label{eq:s_rv}
\sigma_{v_{rad,OC}}=\sqrt{\frac{\sum_i w_i}{(\sum_i w_i)^2+\sum_i w_i^2}\times\sum_i w_i\times(v_{rad,i}-v_{rad,OC})^2}
\end{equation}

Finally, the uncertainty in the average radial velocity, $e_{v_{rad,OC}}$ is obtained as the maximum of the standard error $\frac{\sigma_{v_{rad,OC}}}{\sqrt{N}}$ and $\frac{I}{\sqrt{N}}$ \citep[][]{jasniewiczmajor1988}, where $N$ is the number of star members and $I$ is the internal error of $v_{rad,OC}$ defined as:

\begin{equation}\label{eq:interror}
I=\frac{\sum_i w_i\times v_{scatter,i}}{\sum_i w_i}
\end{equation}

In order to discard stars with discrepant radial velocities, with respect to $v_{rad,OC}$, we applied an iterative $\kappa$-$\sigma$ clipping algorithm removing those objects with velocities outside the range $v_{rad,OC}\pm3\times\sigma_{v_{rad,OC}}$. In this analysis we have also discarded those objects previously reported as spectroscopic binaries or with large $v_{scatter}$ values, which may be a sign of binarity. For those stars observed with more than one telescope, excluding CAH2, we have used the weighted mean and standard deviation obtained using Eqs. \ref{eq:meanrv} and \ref{eq:s_rv}, respectively. The objects observed with CAH2 have been excluded of the analysis except for those clusters where the stars have only been observed with this instrument: NGC\,2126, NGC\,6755, and UBC\,106.

Figures~\ref{fig:rv_membership} and \ref{fig:apex_rv_mermbership} show the determination of the average radial velocities for each cluster. The obtained values are listed in Table~\ref{tab:oc_avglit}. Notes about each particular cluster/star can be found in Appendix~\ref{apex:indvnotes}. For four of the clusters, ASCC\,108, COIN-Gaia\,11, NGC\,6603, and UPK\,55, we have not been able to constrain their average radial velocities in spite of having two or more stars observed in each of them but with significantly different radial velocities (see Appendix~\ref{apex:indvnotes} for details and Fig.~\ref{fig:rv_nomembership}). A single star has been sampled in two clusters: Alessi\,1 and Melotte\,72. Therefore, we provide radial velocities for a total of 47 clusters. To our knowledge this is the first\footnote{Radial velocities for these clusters have been already published by \citet{tarricq2021new} but they used our determinations as cited in that paper.} radial velocity determination from high-resolution spectroscopy, R$\geq$20\,000 for 20 clusters: Alessi\,1, FSR\,0278, FSR\,0850, Melotte\,72, NGC\,559, NGC\,609, NGC\,2126, NGC\,2266, NGC\,6645, NGC\,6728, NGC\,6939, NGC\,6997, NGC\,7245, Ruprecht\,171, Skiff\,J1942+38.6, UBC\,3, UBC\,6, UBC\,44, UBC\,59, and UBC\,215. In fact, for the last five, this is the first-ever radial velocity determination. For the remaining clusters, the derived radial velocities are in agreement with the average radial velocities for these clusters available in the literature.

\subsection{Comparison with the literature}

\begin{table*}
\setlength{\tabcolsep}{1.25mm}
\begin{center}
	\caption{Average radial velocities derived here for the OCs in our sample, together with other radial velocity determinations available in the de literature.}
	\label{tab:oc_avglit}
	\begin{tabular}{lcccccccl} 
		\hline
		Cluster & $v_{\rm rad,OC}$ & $\sigma_{v_{\rm rad,OC}}$ & $e_{v_{\rm rad,OC}}$ & N & $v_{\rm rad,lit}$ & $\sigma_{\rm lit}$ & N$_{\rm lit}$ & Source \\
	     & [\kms] & [\kms] & [\kms] & & [\kms] & [\kms] & &\\
\hline
  Alessi\,1 & -4.67 &  & 0.03 & 1 &  &  &  & \\
  Berkeley\,17 & -73.58 & 0.18 & 0.07 & 6 & -73.4 & 0.4 & 7 & 3\\
  FSR\,0278 & -6.76 & 1.83 & 0.75 & 6 &  &  &  & \\
  FSR\,0850 & 18.37 & 0.26 & 0.13 & 4 &  &  &  & \\
  IC\,4756 & -25.21 & 0.53 & 0.20 & 7 & -25.8 & 0.6 & 13 & \\
  King\,1 & -52.98 & 0.84 & 0.38 & 5 & -53.1 & 3.1 & 28 & 5\\
  Melotte\,111 & -0.10 & 0.37 & 0.12 & 9 & 0.01 & 0.08 & 28 & \\
  Melotte\,72 & 70.70 &  & 0.11 & 1 &  &  &  & \\
  NGC\,188 & -41.64 & 0.56 & 0.25 & 5 & -42.36 & 0.64 & 473 & WOCS\\
  NGC\,559 & -77.72 & 0.29 & 0.20 & 2 &  &  &  & \\
  NGC\,609 & -44.32 & 0.68 & 0.28 & 6 &  &  &  & \\
  NGC\,752 & 5.46 & 0.60 & 0.30 & 4 & 5.54 & 0.14 & 54 & \\
  NGC\,1817 & 65.60 & 0.22 & 0.11 & 4 & 65.31 & 0.09 & 31 & \\
  NGC\,1907 & 2.51 & 0.52 & 0.26 & 4 & 2.3 & 0.5 & 5 & \\
  NGC\,2099 & 8.53 & 0.79 & 0.25 & 10 & 8.3 & 0.2 & 30 & \\
  NGC\,2126 & -11.12 & 0.50 & 0.35 & 2 &  &  &  & \\
  NGC\,2266 & 54.24 & 0.86 & 0.38 & 5 &  &  &  & \\
  NGC\,2354 & 34.22 & 0.30 & 0.15 & 4 & 33.45 & 0.35 & 8 & \\
  NGC\,2355 & 35.46 & 0.29 & 0.13 & 5 & 35.02 & 0.16 & 7 & \\
  NGC\,2420 & 73.61 & 0.15 & 0.07 & 5 & 74.63 & 0.61 & 395 & GES\\
  NGC\,2539 & 29.01 & 0.60 & 0.27 & 5 & 28.89 & 0.21 & 11 & \\
  NGC\,2632 & 28.84 & 0.15 & 0.10 & 2 & 34.76 & 0.07 & 30 & \\
  NGC\,2682 & 33.96 & 0.44 & 0.16 & 8 & 33.64 & 0.85 & 1278 & WOCS\\
  NGC\,6633 & -28.70 & 0.08 & 0.04 & 3 & -28.18 & 0.77 & 34 & GES\\
  NGC\,6645 & -3.05 & 0.77 & 0.31 & 6 &  &  &  & \\
  NGC\,6705 & 35.49 & 0.88 & 0.22 & 16 & 35.53 & 2.3 & 540 & GES\\
  NGC\,6728 & 11.08 & 0.33 & 0.15 & 5 &  &  &  & \\
  NGC\,6755 & 26.90 & 0.90 & 0.45 & 4 & 26.63 & 0.04 & 2 & \\
  NGC\,6791 & -46.49 & 1.39 & 0.52 & 7 & -47.4 & 1.1 & 111 & WOCS\\
  NGC\,6811 & 6.91 & 0.35 & 0.16 & 5 & 6.68 & 0.08 & 5 & \\
  NGC\,6819 & 2.96 & 0.51 & 0.26 & 4 & 2.45 & 1.02 & 679 & WOCS\\
  NGC\,6939 & -18.45 & 0.47 & 0.19 & 6 &  &  &  & \\
  NGC\,6940 & 8.53 & 0.72 & 0.32 & 5 & 7.89 & 0.14 & 21 & \\
  NGC\,6991 & -12.63 & 0.27 & 0.16 & 3 & -21.77 & 0.91 & 13 & \\
  NGC\,6997 & -19.39 & 0.99 & 0.41 & 6 &  &  &  & \\
  NGC\,7142 & -49.71 & 2.83 & 1.26 & 5 &  &  &  & \\
  NGC\,7245 & -75.86 & 3.00 & 1.50 & 4 &  &  &  & \\
  NGC\,7762 & -46.63 & 0.88 & 0.39 & 5 &  &  &  & \\
  NGC\,7789 & -53.51 & 0.25 & 0.13 & 4 & -53.5 & 1.5 & 564 & WOCS\\
  Ruprecht\,171 & 6.18 & 0.29 & 0.12 & 6 &  &  &  & \\
  Skiff\,J1942+38.6 & -18.53 & 0.26 & 0.11 & 5 &  &  &  & \\
  UBC\,3 & 2.45 & 0.75 & 0.37 & 4 &  &  &  & \\
  UBC\,6 & -32.79 & 0.59 & 0.24 & 6 &  &  &  & \\
  UBC\,44 & -38.40 & 0.07 & 0.05 & 2 &  &  &  & \\
  UBC\,59 & -35.79 & 0.074 & 0.04 & 3 &  &  &  & \\
  UBC\,106 & 42.32 & 1.20 & 0.85 & 2 & 41.7 & 1.2 & 9 & Ne\\
  UBC\,215 & 0.34 & 0.30 & 0.15 & 4 &  &  &  & \\
		\hline
	\end{tabular}
\end{center}
\end{table*}

We have compared the average radial velocities obtained for each cluster with the values available in the literature. Several catalogues based on homogeneous measurements have been released in the last years. These are the works performed by \citet{conrad_rave_OC}, \citet{soubiran2018gaiadr2_rv}, and \citet{donor2020dr16} based on RAVE, \gaia DR2, and APOGEE DR16 radial velocities, respectively. In the case of APOGEE we added two clusters, King\,1 and NGC\,1817, from \citet{carrera2019apo} which are not included in the \citet{donor2020dr16} sample, although these determinations are based on a single object in each case. We also compared our results with the values recently published by \citet{tarricq2021new} who have compiled radial velocities from nearly 25\,000 OC members. This sample includes some OCCASO data presented in this paper although without the detailed membership selection discussed in Appendix~\ref{apex:indvnotes}. Finally, we have performed a compilation of other values available in the literature. To do that we have used as a starting point the \citet{kharchenko2013_OC} compilation. We have updated this compilation with the values obtained in the framework of WOCS: NGC\,188, NGC\,2682, NGC\,6791, NGC\,6819, and NGC\,7789. We have done the same with the three clusters which average radial velocities have been obtained by \citet{jackson2020_GESOC} from GES data: NGC\,2420, NGC\,6633, and NGC\,6705. Recently, \citet{spina2021_galahOC} has compiled an OC sample based on APOGEE and GALAH surveys. However, they did not publish average radial velocities for the studied clusters.

\begin{figure}
\centering
\includegraphics[width=\columnwidth]{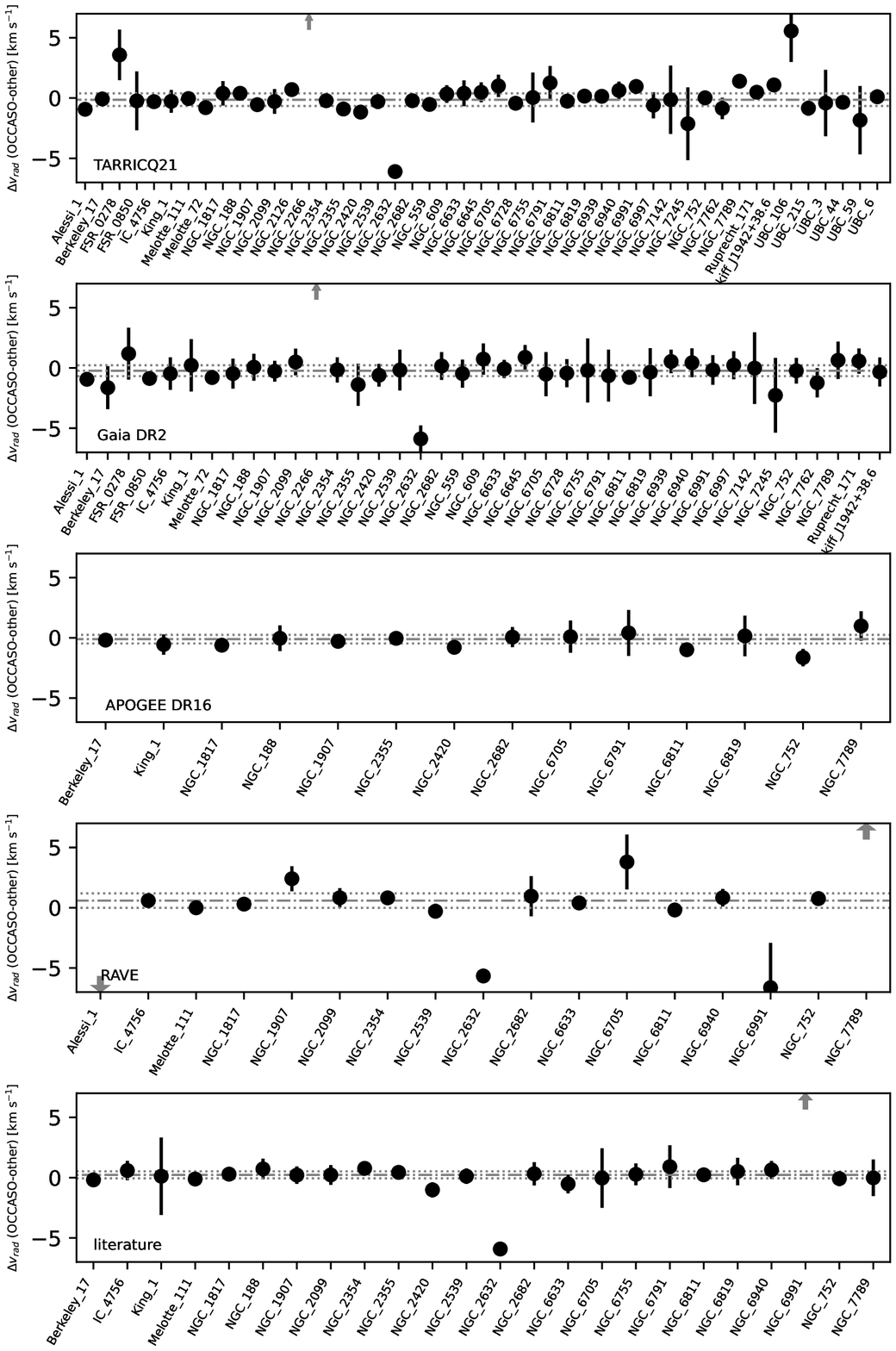}
\caption{Comparison of the average cluster radial velocities with different values available in the literature for each cluster. Arrows denoted that the value is outside the panel range. Dashed and dotted lines show the median and the MAD of the differences as listed in Table~\ref{tab:rv_oc_comp}, respectively.}
\label{fig:rv_oc_comparison}%
\end{figure}

Figure~\ref{fig:rv_oc_comparison} shows the comparison of the average radial velocities of the OCCASO clusters with the samples available in the literature described above. In general, there is a good agreement, within the uncertainties, as listed in Table~\ref{tab:rv_oc_comp}. The average median differences are in a range of $\pm$0.2\kms, and therefore, within the expected uncertainties with median absolute deviation between $\sim$0.3 and $\sim$0.5\kms. The largest values are found in the comparison with \citet{conrad_rave_OC} but that can be explained by the large uncertainties of RAVE radial velocities obtained with a lower spectral resolution.

For some surveys the average differences found here differs from the values found in the star-by-star comparison performed in the previous section, although they are within the sampling errors. The largest differences can be explained by the different membership selection performed in each case, and therefore, the average value is obtained from different stars and/or different number of stars. 

The Praesepe cluster, NGC\,2632, is the one that shows discrepancies with all the values available in the literature. We have observed seven stars in this system and five of them have been reported as spectroscopic binaries in the literature as discussed in Appendix~\ref{apex:indvnotes}. As a consequence, the average value for the cluster is based only on two stars. At least one of them,  \gaia EDR3\,661311443306610688, has a large $v_{err}$ value which could be a sign of rotation. 

NGC\,2266 has discrepant values in comparison with \citet{tarricq2021new} and \citet{soubiran2018gaiadr2_rv} values. The \citet{tarricq2021new} and \citet{soubiran2018gaiadr2_rv} radial velocity of NGC\,2266 has been obtained from only one star. This star is the same used in both cases. Our radial velocity determination for this cluster is based on 5 stars which radial velocities show a good agreement within the uncertainties (see Fig.~\ref{fig:rv_membership}).

Two clusters, FSR\,278 and UBC\,106, show significant differences with the values reported by \citet{tarricq2021new}. Although the 6 stars studied in FSR\,278 show a significant radial velocity dispersion this does not explain the $\sim$3.6\,\kms~ difference with \citet{tarricq2021new} based on 7 stars. The other radial velocity available in the literature for this cluster by \citet{soubiran2018gaiadr2_rv} from \gaia DR2 based on 5 stars shows a good agreement, well within the uncertainties, with the value obtained here. Our UBC\,106 radial velocity is based on only two stars observed with CAH2, which as already mentioned, implies larger uncertainties. On contrary, the \citet{tarricq2021new} determination is based on 9 stars.

For NGC\,6991 we found discrepant values with RAVE and the literature radial velocities but not with the \gaia DR2 one. Because our value is in very good agreement with \gaia DR2, which is based on 78 stars, we assume that our determination is reliable. 

Finally, Alessi\,1 and NGC\,7789 RAVE radial velocities differ significantly from ours and other measurements for these clusters. In the case of Alessi\,1 we have sampled only one star and its radial velocity is in good agreement with other determinations available in the literature. NGC\,7789 is a well-studied cluster, so the RAVE determination for this cluster is suspicious.

\begin{table}
\setlength{\tabcolsep}{1.25mm}
\begin{center}
	\caption{Statistics of the differences in radial velocities in the sense OCCASO-others.}
	\label{tab:rv_oc_comp}
	\begin{tabular}{lccc} 
		\hline
		Survey & Clusters & Median & MAD \\
	 & & [\kms] & [\kms]\\
	 \hline
		\citet{tarricq2021new} & 47 & -0.07 & 0.47 \\
		\citet{soubiran2018gaiadr2_rv} & 39 & -0.20 & 0.43 \\
		\citet{donor2020dr16} & 14 & -0.11 & 0.35 \\
		\citet{conrad_rave_OC} & 17 & 0.59 & 0.60\\
        Literature & 24 & 0.23 & 0.29\\
		\hline
	\end{tabular}
\end{center}
\end{table}

\section{Open clusters kinematics}\label{sec:kinematic}

In order to investigate the kinematics of the observed clusters we study the line-of-sight velocity of the OCs within the context of the Galactic disc, and also coupled with proper motions, distances and ages listed in Table~\ref{tab:ocs}. As explained above, we compute average proper motions from \gaia EDR3 but using the membership probabilities compiled by \citet{cantatgaudin2020}. Distances and ages are also taken from the same authors \citep{cantatgaudin2020} from \gaia DR2 through a machine learning method using both photometry and parallaxes. For the farthest clusters, the addition of photometry allows a better estimation of the distances.  

\subsection{Radial velocities with respect to the GSR and RSR}\label{sec:gsr_rsr}

We compute the line-of-sight velocity with respect to the Galactocentric Standard of Rest (GSR) and with respect to the Regional Standard of Rest (RSR) as done in Paper\,I using:

\begin{align}
\begin{split}
v_\text{GSR}=  &v_\text{rad,OC}+U_{\odot}\cos l \cos b \\
     &+ \left( \Theta_{\text{0}}+V_{\odot} \right) \sin l \cos b + W_{\odot} \sin b
\end{split}
\end{align}
\begin{align}
 v_\text{RSR}=v_\text{GSR}-  \Theta_R \frac{R_0}{R} \sin l \cos b
\end{align}

\noindent where $v_{rad,OC}$ is the average OC heliocentric radial velocity derived in the previous section;  ($U_{\odot}$, $V_{\odot}$, $W_{\odot}$)
are the components of the motion of the Sun with respect to the Local Standard of Rest (LSR); $\Theta_0$ and $\Theta_R$ are the circular velocities
at the Galactocentric distances of the Sun $R_0$, and the cluster $R$, respectively.
For the Sun, we adopted $(U_{\odot},V_{\odot}, W_{\odot})= (11.1, 12.24, 7.25)$\,\kms~from \citet{Schonrich}, and $R_0=8.34$\,kpc from \citet{Reid}. For the circular velocity around the Galactic centre, $\Theta_{\text{0}}$ is adopted as 240\kms~from \citet{Reid} and $\Theta_{\text{R}}$ is computed according to the Galactic potential described in Sec.~\ref{sec:orbits}.

The results are listed in Table~\ref{tab:GSR_RSR}. The uncertainties have been estimated with 100\,000 realizations taking into account the errors in radial velocities and distances of the clusters. All clusters show $v_\text{RSR}$ values typical of the thin disc kinematics, with a median value of $-1.3\pm13.5$\,\kms. Berkeley\,17, the oldest and farthest OC in our sample, is the only cluster showing a large $v_\text{RSR}$ value of $-78.5\pm0.4$\,\kms. 

\begin{table*}
\setlength{\tabcolsep}{1.25mm}
\begin{center}
	\caption{Line-of-sight velocity $v_\text{GSR}$ with respect to the Galactocentric Standard of Rest and $v_\text{RSR}$ with respect to the Regional Standard of Rest. $(U_\text{s},V_\text{s},W_\text{s})$ are the components of the spatial velocity with respect to the Regional Standard of Rest. $V_{\phi}$ is the total azimuthal velocity of the cluster.}
	\label{tab:GSR_RSR}
	\begin{tabular}{lcccccc} 
		\hline
Cluster &  $v_\text{GSR}$  &   $v_\text{RSR}$ &   $U_\text{s}$      &   $V_\text{s}$ & $W_\text{s}$  & $V_{\phi}$\\
 & [\kms] & [\kms] & [\kms] & [\kms] &[\kms] & [\kms]\\
\hline
Alessi\,1          & 192.9$\pm$1.0 &   6.9$\pm$1.3 & -16.8$\pm$2.8 &  -7.093$\pm$1.77 & -12.070$\pm$2.02 & 231.9$\pm$1.8\\
Berkeley\,17       & -66.0$\pm$0.2 & -78.5$\pm$0.4 &  75.5$\pm$0.7 & -10.0$\pm$2.8 &  42.0$\pm$3.6 & 220.8$\pm$3.3\\
FSR\,0278          & 244.8$\pm$1.8 &   9.9$\pm$2.0 &  46.1$\pm$3.3 &  19.1$\pm$3.1 &  -9.4$\pm$1.8 & 259.0$\pm$3.0\\
FSR\,0850          &  -8.4$\pm$0.3 &   2.8$\pm$0.4 &  -2.4$\pm$0.4 & -12.7$\pm$2.6 &   3.9$\pm$1.3 & 221.0$\pm$3.1\\
IC\,4756           & 133.1$\pm$0.5 & -16.4$\pm$0.9 & -11.7$\pm$0.6 & -11.3$\pm$1.0 &  -3.3$\pm$1.0 & 230.0$\pm$0.9\\
King\,1            & 160.6$\pm$0.8 & -23.5$\pm$2.4 &  35.6$\pm$1.0 &  -1.3$\pm$3.4 &  -0.5$\pm$1.3 & 236.0$\pm$3.1\\
Melotte\,111       & -13.3$\pm$0.4 &   5.3$\pm$0.4 &   8.9$\pm$0.3 &   6.5$\pm$0.6 &   6.5$\pm$0.4 & 246.7$\pm$0.6\\
NGC\,188           & 151.5$\pm$0.6 & -15.5$\pm$1.9 &  12.5$\pm$1.9 &  -0.4$\pm$2.0 & -16.0$\pm$1.3 & 237.0$\pm$1.8\\
NGC\,559           & 116.6$\pm$0.3 & -33.8$\pm$3.4 &  51.6$\pm$1.1 &   9.4$\pm$5.8 &  -0.1$\pm$1.7 & 243.6$\pm$5.2\\
NGC\,609           & 148.5$\pm$0.7 &  23.3$\pm$4.6 & -27.1$\pm$4.2 &   1.1$\pm$2.0 &  -0.8$\pm$1.8 & 231.0$\pm$1.6\\
NGC\,752           & 153.3$\pm$0.6 &   8.5$\pm$0.8 & -14.6$\pm$2.0 & -10.6$\pm$2.6 & -13.4$\pm$1.8 & 228.6$\pm$2.7\\
NGC\,1817          &  26.7$\pm$0.2 &  47.2$\pm$0.4 & -46.7$\pm$0.5 &   2.6$\pm$0.7 &  -9.0$\pm$0.7 & 237.6$\pm$1.0\\
NGC\,1907          &  23.9$\pm$0.5 &  -1.4$\pm$0.7 &   0.9$\pm$0.8 &  -4.0$\pm$1.8 &  -8.4$\pm$1.7 & 231.4$\pm$2.2\\
NGC\,2099          &   8.2$\pm$0.8 &  -0.1$\pm$0.8 &  -0.8$\pm$0.8 & -23.0$\pm$3.5 &  -0.8$\pm$1.5 & 212.9$\pm$3.9\\
NGC\,2126          &  51.1$\pm$0.5 &  -7.1$\pm$0.9 &   7.0$\pm$1.3 &  -2.9$\pm$1.4 &   2.0$\pm$0.8 & 233.7$\pm$1.7\\
NGC\,2266          &  11.1$\pm$0.9 &  33.5$\pm$1.1 & -33.8$\pm$1.0 &   0.5$\pm$1.6 &   1.8$\pm$1.9 & 231.7$\pm$2.0\\
NGC\,2354          &-185.7$\pm$0.3 &  -1.9$\pm$1.8 &   7.7$\pm$1.4 &  -2.3$\pm$1.5 &  -7.9$\pm$1.3 & 235.5$\pm$1.3\\
NGC\,2355          & -70.9$\pm$0.3 &   4.6$\pm$1.4 & -11.9$\pm$0.7 &   8.0$\pm$1.4 & -21.0$\pm$3.5 & 242.9$\pm$1.1\\
NGC\,2420          &  -7.7$\pm$0.1 &  46.5$\pm$1.2 & -41.7$\pm$0.9 & -18.1$\pm$1.5 &   9.8$\pm$2.3 & 215.2$\pm$1.9\\
NGC\,2539          &-175.6$\pm$0.6 &  -3.0$\pm$1.7 &  14.0$\pm$1.8 &  -8.3$\pm$0.7 &  -0.2$\pm$1.4 & 229.6$\pm$0.7\\
NGC\,2632          & -68.9$\pm$0.1 &  18.5$\pm$0.2 & -25.2$\pm$1.5 &  -5.5$\pm$1.0 &  -5.5$\pm$2.8 & 234.3$\pm$1.0\\
NGC\,2682          & -94.7$\pm$0.4 &  16.4$\pm$0.8 & -26.3$\pm$1.6 &  -9.3$\pm$0.9 & -15.2$\pm$3.9 & 229.0$\pm$0.9\\
NGC\,6633          & 128.3$\pm$0.1 & -18.2$\pm$0.6 & -17.4$\pm$0.6 &  -7.5$\pm$0.5 &  -0.9$\pm$0.7 & 233.6$\pm$0.5\\
NGC\,6645          &  75.5$\pm$0.8  &  -8.1$\pm$2.3 & -11.6$\pm$2.5 &   7.0$\pm$1.0 &  -5.4$\pm$1.5 & 251.4$\pm$0.9\\
NGC\,6705          & 160.5$\pm$0.9 &  15.5$\pm$4.1 &  27.0$\pm$2.2 & -10.5$\pm$3.6 &   0.1$\pm$1.5 & 234.2$\pm$3.3\\
NGC\,6728          & 129.4$\pm$0.3 &  -1.3$\pm$3.0 &  -6.3$\pm$3.2 &   5.1$\pm$1.4 & -11.1$\pm$1.8 & 249.3$\pm$1.2\\
NGC\,6755          & 192.5$\pm$0.9 &  -4.0$\pm$4.9 &  -2.8$\pm$4.5 &  -3.2$\pm$3.1 &  -5.8$\pm$1.7 & 241.5$\pm$2.8\\
NGC\,6791          & 191.3$\pm$1.4 & -41.7$\pm$2.6 & -63.6$\pm$5.6 & -51.7$\pm$2.5 & -14.0$\pm$2.6 & 189.5$\pm$2.6\\
NGC\,6811          & 252.8$\pm$0.3 &  17.9$\pm$0.4 &  28.8$\pm$1.7 &  16.1$\pm$0.5 &   3.2$\pm$0.9 & 256.7$\pm$0.5\\
NGC\,6819          & 246.8$\pm$0.5 &   9.1$\pm$0.8 &  -9.8$\pm$2.5 &   6.0$\pm$0.6 &  16.6$\pm$1.7 & 247.0$\pm$0.6\\
NGC\,6939          & 227.1$\pm$0.5 &   4.3$\pm$1.5 &   9.7$\pm$1.0 &  10.3$\pm$2.1 & -10.2$\pm$1.6 & 249.4$\pm$1.9\\
NGC\,6940          & 246.4$\pm$0.7 &  13.2$\pm$1.0 &  24.9$\pm$1.4 &   6.1$\pm$1.2 & -14.7$\pm$2.1 & 247.1$\pm$12.1\\
NGC\,6991          & 240.0$\pm$0.3 &  -0.1$\pm$0.3 & -34.0$\pm$4.3 &  -1.2$\pm$0.4 &   9.9$\pm$1.0 & 239.0$\pm$0.4\\
NGC\,6997          & 232.9$\pm$1.0 &  -7.2$\pm$1.0 &  18.9$\pm$1.1 &  -6.6$\pm$1.0 &   1.9$\pm$0.8 & 233.6$\pm$1.0\\
NGC\,7142          & 188.5$\pm$2.8 & -15.1$\pm$3.9 &   2.1$\pm$2.5 & -17.5$\pm$3.1 &   7.3$\pm$1.5 & 220.0$\pm$2.9\\
NGC\,7245          & 168.9$\pm$3.0 & -34.3$\pm$4.8 &  32.7$\pm$1.9 & -20.2$\pm$5.8 &   1.8$\pm$1.4 & 216.5$\pm$5.4\\
NGC\,7762          & 172.2$\pm$0.9 & -28.4$\pm$1.4 &   5.2$\pm$2.5 & -32.6$\pm$0.9 &  17.5$\pm$1.6 & 206.3$\pm$1.0\\
NGC\,7789          & 167.7$\pm$0.2 & -20.5$\pm$2.7 &   3.7$\pm$2.7 & -23.5$\pm$1.3 &  -5.2$\pm$2.1 & 213.4$\pm$1.0\\
Ruprecht\,171      &  87.7$\pm$0.3 &   4.2$\pm$1.8 & -12.6$\pm$3.4 &  40.6$\pm$2.6 & -39.1$\pm$4.5 & 284.4$\pm$2.9\\
Skiff\,J1942+38.6  & 224.4$\pm$0.3 & -14.4$\pm$0.3 &  14.5$\pm$1.1 & -15.3$\pm$0.5 &   3.4$\pm$0.6 & 225.9$\pm$0.5\\
UBC\,3             & 186.7$\pm$0.7 &  -9.0$\pm$2.8 & -19.4$\pm$3.6 &   2.2$\pm$1.5 &   6.4$\pm$0.6 & 245.3$\pm$1.3\\
UBC\,6             & 204.6$\pm$0.6 & -10.7$\pm$1.7 &  25.4$\pm$0.7 &  -1.0$\pm$2.2 &  -6.0$\pm$1.8 & 237.7$\pm$2.0\\
UBC\,44            & 134.7$\pm$0.1 &  -0.5$\pm$3.0 &  11.2$\pm$2.2 &  15.3$\pm$3.2 &   0.2$\pm$2.4 & 249.4$\pm$2.7\\
UBC\,59            &  31.7$\pm$0.1 & -23.3$\pm$1.3 &  20.2$\pm$1.9 & -13.1$\pm$2.6 &  -4.2$\pm$2.3 & 219.9$\pm$3.0\\
UBC\,106           & 166.8$\pm$1.2 &  19.8$\pm$4.6 &  13.1$\pm$4.7 &  15.6$\pm$1.9 &  10.5$\pm$1.0 & 260.6$\pm$1.6\\
UBC\,215           &-158.7$\pm$0.3 & -35.7$\pm$1.7 &  43.6$\pm$2.7 &  -2.0$\pm$1.5 &  -3.1$\pm$1.1 & 234.7$\pm$1.8\\
\hline
\end{tabular}
\end{center}
\end{table*}

\subsection{Spatial velocity with respect to GSR and RSR}\label{sec:spatial_vel_gsr_rsr}

Our radial velocities have been combined with the mean proper motions of the clusters using \gaia EDR3 to derive full spatial velocities with respect to GSR $(V_\text{R}, V_{\phi}, V_\text{z})$ and to RSR $(U_\text{s},V_\text{s},W_\text{s})$, being $U_\text{s}=-V_\text{R}$, $V_\text{s}=V_\phi-\Theta_R$ and
$W_\text{s}=V_\text{z}$. 

\begin{figure}
\centering
\includegraphics[width=\columnwidth]{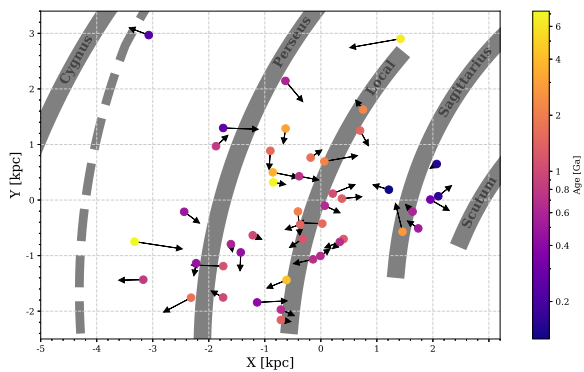}
\caption{Projection on the Galactic plane of the position and velocity with respect to the Regional Standard of Rest of the clusters in our sample.}
\label{fig:ux_vy}
\end{figure}

These values are also included in Table~\ref{tab:GSR_RSR}. Uncertainties have decreased significantly with respect to Paper\,I because of the huge improvement of proper motions with respect to the pre-\gaia~era. Median and MAD values of $(U_\text{s},V_\text{s},W_\text{s})$ are $(2.9\pm16.7,-2.6\pm8.3,-0.9\pm6.1)$\,\kms. The clusters with the largest $v_{RSR}$ values are Berkeley\,17, NGC\,6791 and Ruprecht\,171, but still in the range $-63$ to $+75$\,\kms.
Figure~\ref{fig:ux_vy} shows the projection on the Galactic plane of the position and velocity with respect to the RSR of the clusters in our sample. 

In Paper\,I we pointed out the similarity of ages and non-circular velocities of IC~4756 and NGC~6633 both in the Local arm and close together. This similarity in velocity is confirmed with the new proper motions of \gaia and our new radial velocities. In the pre-\gaia era, the log age determination of both clusters was in the range $8.6-8.7$\,dex. However, according to \citet{cantatgaudin2020}, IC\,4756 and NGC\,6633 have log ages of 9.1 and 8.8\,dex, respectively. \citet{Dias2021} also gives log ages of IC\,4756 and NGC~6633 of 9.0 and 8.8\,dex, respectively. Both papers indicate a $0.2-0.3$\,dex difference in log age between both clusters. The study of their birthplace (see section \ref{sec:orbits}) neither indicates a common origin. Most likely, a common origin can be discarded due to the age difference.

\subsection{Open clusters orbits}
\label{sec:orbits} 

To complete our analysis we have integrated the orbits of the OCs in our sample. Due to the uncertainty in the determination of the real Galactic potential, we consider three different Galactic models proposed in the literature. The first one proposed by \citet{Bovy2015}, named \textit{ MW2014}, is an axisymmetric potential composed of a spherical bulge, a Miyamoto-Nagai disc and a halo with a Navarro-Frenk-White profile \citep[NFW,][]{nfw97}. 
The second and third models are based on the previous one but adding two non-axisymmetric components. For the second model we add a bar characterized as a Ferrers potential \citep{Ferrers1877} with n=2, the semi-major, middle and minor axes are fixed to 3\,kpc, 0.35\,kpc and 0.2375\,kpc, respectively, the bar mass is $10^{10}\,M_{\odot}$ \citep{Romero-Gomez2015} and a constant pattern speed fixed to $\Omega=42\,$\,km\,s$^{-1}$\,kpc$^{-1}$ \citep{Bovy2019}, which puts co-rotation at $R=5.6$\,kpc and the Outer Lindblad Resonance at $R=9.$\,kpc. The angular orientation of the bar with respect to the Sun-Galactic Centre line is $20^{\circ}$ \citep[][and references therein]{Romero-Gomez2011}. The third model adds a sinusoidal spiral arms potential from \citet{Cox-Gomez2002}. We model two spiral arms with an amplitude of $0.4$ and a pattern speed of $\Omega=21$\,km\,s$^{-1}$\,kpc$^{-1}$ \citep[e.g.][]{Antoja2011}, which puts co-rotation at $R=10.6$\,kpc. 

Using the python {\sl galpy} package \citep{Bovy2015}, we have integrated the orbit backwards in time during the age of the cluster with a step of 2\,Ma. The components of the motion of the Sun with respect to the LSR, the Galactocentric distance of the Sun, the circular velocity at this distance and the distances and ages of the OCs are the ones described in Sect.~\ref{sec:gsr_rsr}. It is important to bear in mind that for the older clusters, therefore larger integrated times, the derived orbits are more uncertain because of the temporal evolution of the potential and the lack of knowledge about the interactions of the clusters with the disc structures such as molecular clouds or spiral arms along the time.

\begin{figure}
\centering
\includegraphics[width=\columnwidth]{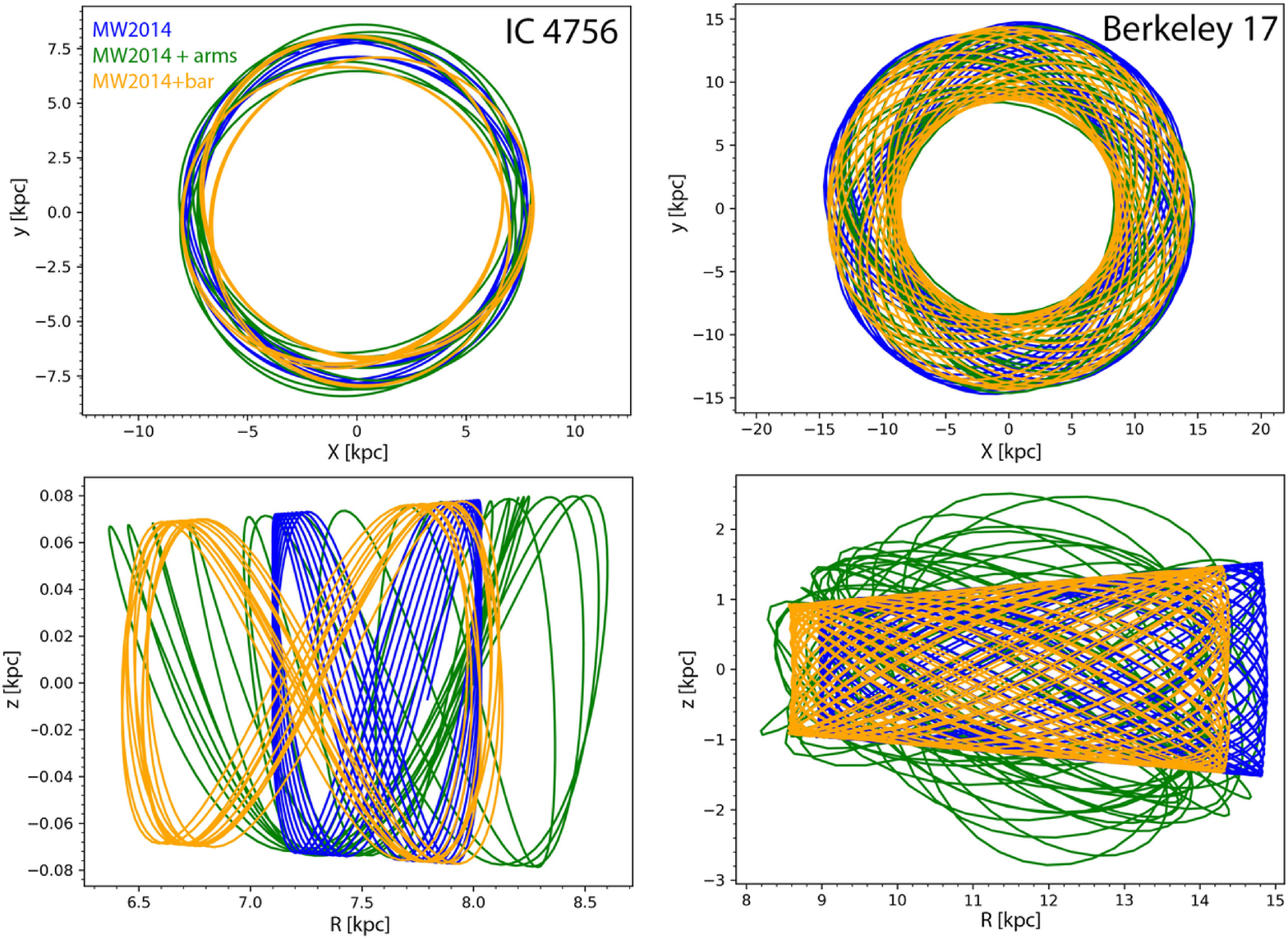}
\caption{Example of the orbits for IC\,4756 with an age of 1.29\,Ga (left) and Berkeley\,17 with 7.24\,Ga (right)}
\label{fig:orb_exp}
\end{figure}

We have also verified that using the distances and ages provided by \citet{Dias2021} does not significantly change our results. In order to calculate the uncertainties of the parameters, we have carried out a Monte Carlo sampling of the radial velocities, proper motions, distances and their uncertainties, which we assume as Gaussian. We have taken 100 realizations of values from the Monte Carlo sampling and integrated the orbits, considering the standard deviation of the calculated parameters as uncertainties. As an example, the orbits derived using the three potentials in different planes for two of the clusters in our sample is shown in Fig.~\ref{fig:orb_exp}.

\begin{figure}
\centering
\includegraphics[width=\columnwidth]{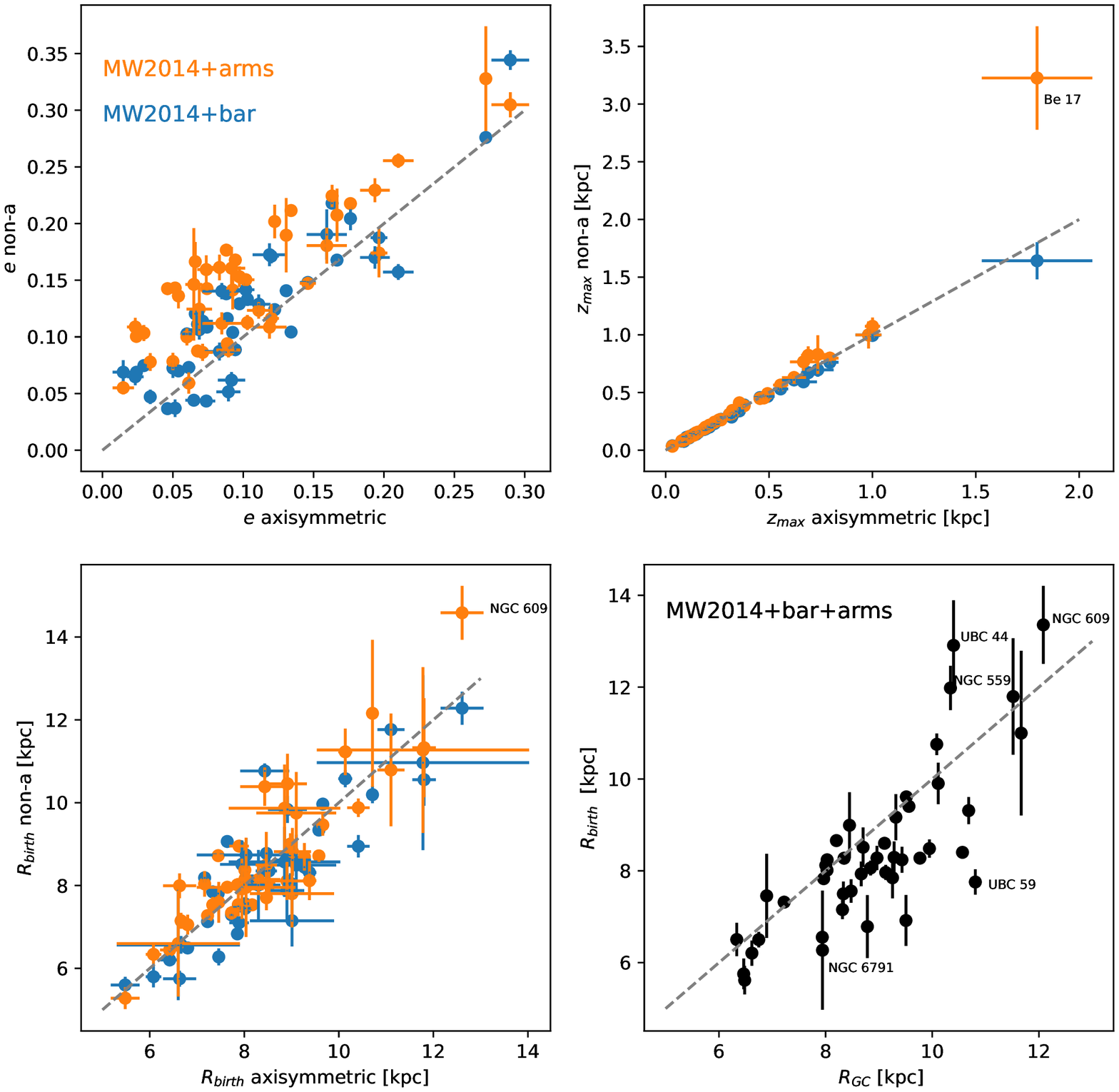}
\caption{Comparison between the results obtained by integrating the orbits with the \textit{MW2014} potential, and adding the arms (in orange) and bar (in blue) potentials for the eccentricity (top left), the maximum height with respect to the galactic plane (top right), the birth radius (bottom left). And the comparison between the current Galactocentric radius and the calculated birth radius for \textit{MW2014} adding the bar and the spiral potential together (bottom right).}
\label{fig:comp}
\end{figure}

In Fig.~\ref{fig:comp}, we study the effect of the bar and spiral models on the eccentricity, $e$ (top left), maximum height with respect to the Galactic plane, $z_{max}$ (top right). Regarding $z_{max}$, it barely varies among the three potentials, being the median $z_{max}$ 0.22\,kpc for all Galactic models, except in the case of Berkeley\,17 where the addition of spiral arms causes the orbit to increase its vertical excursion up to 3.23\,kpc (see bottom right panel of Fig.~\ref{fig:orb_exp}).

The addition of both bar or spiral arms yield an increase of eccentricity in most of the cases. The median value of the eccentricity calculated with the \textit{MW2014} potential is 0.08, while when adding the bar the median becomes 0.11, and with the spiral arms the median increases to 0.14, which means an increase of the 37.5\,\% and 75\,\%, respectively.

The birth radius is also modified as a function of the potential used (see Fig.~\ref{fig:comp} bottom left).
The birth radius of clusters with ages greater than 1~Ga is more uncertain and those older clusters give as well the greater differences depending on the choice of potential.
The exception is the NGC\,609 cluster at 220\,Ma old. 
In the bottom right panel of Fig.~\ref{fig:comp} we represent the birth radius with respect to its current position. The potential used in this case is the sum of the axisymmetric \textit{MW2014}, bar and spiral potentials. About 70\,\% of the OCs of our sample have been formed in the innermost regions and migrated outward. We find that the outward migrated systems are the ones with current $R_{GC}$ below 11\,kpc.
The clusters UBC\,59 and NGC\,6791 stand out with a migration of 3.05 and 1.67\,kpc respectively. The outer clusters NGC\,609, NGC\,559 and UBC\,44 show the opposite trend. They appear to have been born in outer regions of the galaxy than their current radius.

\begin{figure}
\centering
\includegraphics[width=\columnwidth]{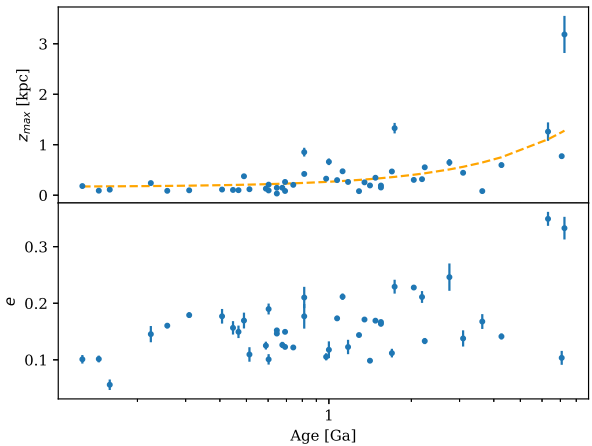}
\caption{The run of $z_{max}$ (top) and eccentricity (bottom) as a function of age for the OCs in our sample using the \textit{MW2014} adding bar and spiral components.}
\label{fig:zmax_e-age}
\end{figure}

Figure~\ref{fig:zmax_e-age} shows the run of $z_{max}$ (top) and eccentricity (bottom) as a function of age for the studied clusters. The orbits have been computed with the potential \textit{MW2014} adding bar and spiral components. Similar results are obtained using only the potential \textit{MW2014} and including the spiral and bar potential separately (see Appendix~\ref{apex:zmax_e}). Regardless of the chosen potential, there is a clear dependency of $z_{max}$ with age, as it was also reported by \citet{tarricq2021new}.
This may be explained by the survival bias of OCs and the dynamical heating of the thin disc. Open Clusters are destroyed due to the interaction with the spiral arms, the bar, and the giant molecular clouds that the disc contains \citep{spitzer1951,jenkins1990}. Therefore, the clusters that have larger $z_{max}$ values spend more time away from the destructive influence of the disc. Hence, surviving old clusters may be the ones with the highest $z_{max}$. On the other hand, the dynamical heating of the disc implies an increase of $z_{max}$ with time. It is worth to mention the case of King\,1, a 3.9\,Ga old cluster which orbit is very close to the galactic plane, with a $z_{max}$ of 80\,pc. A possible explanation for the survival of this cluster is that it was much more massive at the moment of birth. Our sample does not contain other examples of OCs older than 2\,Ga with low $z_{max}$ but larger samples point towards the existence of other systems with these features \citep[e.g.][]{tarricq2021new}.

Regarding the eccentricity, we see that the dispersion of the eccentricity increases as a function of age, in line with the results obtained by \citet{tarricq2021new}. This is due to the previously mentioned effects, the older clusters are more likely to have interacted with non-axisymmetric components of the Galactic potential and giant molecular clouds, increasing the eccentricity of their orbits.

\label{sec:orbits}

\section{Summary}\label{sec:conclusion}

The OCCASO survey aims to complement the massive Galactic spectroscopic surveys by obtaining high-resolution spectra of open clusters. At the moment, OCCASO has completed more than 130 observing nights, in which spectra for a total of 336 stars have been acquired: 312 objects belonging to 51 clusters and 24 \gaia Benchmark Stars. In this paper we have revised the observational strategy now mainly based on \gaia data. A new, completely automatic, data reduction pipeline has been presented, yielding a significant improvement in the quality of the output spectra. 

Radial velocities have been obtained for the sample stars using cross-correlation with a library of synthetic spectra which covers from early M to A spectral types. The typical internal uncertainties, determined from the scatter of the individual measurements of each star $v_{scatter}$, go from 10\,m\,s$^{-1}$ for MERC and NOT1 to 21.2\,m\,s$^{-1}$ for CAH2. Radial velocities derived from MERC, NOT1, and NOT2 instrumental configurations are compatible within the uncertainties. Velocities derived CAH2 show a larger scatter, but always below 0.5\,\kms.

The derived radial velocities together with the \gaia proper motions and parallaxes have been used to investigate the membership of the observed stars and to derive the average radial velocities for a total of 47 clusters. This is the first radial velocity determination from high-resolution spectra for about 20 systems and the first-ever determination for five of them: UBC\,3, UBC\,6, UBC\,44, UBC\,59, and UBC\,215. With the information already on hand, we cannot ensure that we have sampled a real cluster member for four systems: ASCC\,108, COIN-Gaia\,11, NGC\,6603, and UPK\,55. 

The obtained radial velocities together with average \gaia proper motions, distances and ages have been used to investigate the kinematics of the sampled clusters. They mainly follow the Galactic thin disc kinematics. With the new precision of our results we can see that the clusters IC\,4756 and NGC\,6633 indeed have similar non-circular velocities, as pointed out in Paper\,I, although their ages are not compatible, so their common origin can be discarded.  

Finally, we have integrated their orbits using  three different Galactic potential models. The effect of including the bar and spiral arms is not important regarding the height above the plane except for the oldest cluster Berkeley\,17, but it does increase the eccentricity in most of the cases: a median of 37.5\% adding the bar and a 75\% with the spiral arms. About 70\% of the OCs of our sample have been formed in the innermost regions and migrated outward, while the three outer clusters (NGC\,609, NGC\,559 and UBC\,44) were born in outer regions than their current radius. In all the models, the height above the plane and the dispersion of the eccentricity increases as a function with age. 

\begin{acknowledgements}

We acknowledge the anonymous referee for his/her help in making the paper more readable. We also acknowledge C. Allende-Prieto for useful discussions and comments about this work. This work has made use of data from the European Space Agency (ESA) mission \gaia (\url{https://www.cosmos.esa.int/gaia}), processed by the \gaia Data Processing and Analysis Consortium (DPAC, \url{https://www.cosmos.esa.int/web/gaia/dpac/consortium}). Funding for the DPAC has been provided by national institutions, in particular the institutions participating in the \gaia Multilateral Agreement.

Based on observations made with the Nordic Optical Telescope, owned in collaboration by the University of Turku and Aarhus University, and operated jointly by Aarhus University, the University of Turku and the University of Oslo, representing Denmark, Finland and Norway, the University of Iceland and Stockholm University at the Observatorio del Roque de los Muchachos, La Palma, Spain, of the Instituto de Astrof\'{\i}sica de Canarias. Based on observations made with the Mercator Telescope, operated on the island of La Palma by the Flemish Community, at the Spanish Observatorio del Roque de los Muchachos of the Instituto de Astrof\'{\i}sica de Canarias. Based on observations obtained with the HERMES spectrograph, which is supported by the Research Foundation - Flanders (FWO), Belgium, the Research Council of KU Leuven, Belgium, the Fonds National de la Recherche Scientifique (F.R.S.-FNRS), Belgium, the Royal Observatory of Belgium, the Observatoire de Genève, Switzerland and the Thüringer Landessternwarte Tautenburg, Germany. Based on observations collected at Centro Astron\'omico Hispano en Andaluc\'{\i}a (CAHA) at Calar Alto, operated jointly by Instituto de Astrof\'{\i}sica de Andaluc\'{\i}a (CSIC) and Junta de Andaluc\'{\i}a.

This research has made use of NASA's Astrophysics Data System Bibliographic Services.

This research made use of Astropy,\footnote{http://www.astropy.org}  \citep{astropy:2013, astropy:2018}, Matplotlib \citep[][]{matplotlib}, Sk-learn \citep{scikit-learn} python packages, TopCat \citep{Taylor2005} and galpy\footnote{http://github.com/jobovy/galpy} \citep{Bovy2015} the IDL software.

This work was partially supported by the Spanish Ministry of Science, Innovation and University (MICIU/FEDER, UE) through grant RTI2018-095076-B-C21, and the Institute of Cosmos Sciences University of Barcelona (ICCUB, Unidad de Excelencia ’Mar\'{\i}a de Maeztu’) through grant CEX2019-000918-M. J.L-B. acknowledges financial support received from ”la Caixa” Foundation (ID 100010434) and from the European Union’s Horizon 2020 research and innovation programme under the Marie Skłodowska-Curie grant agreement No 847648, with fellowship code LCF/BQ/PI20/11760023. This research has also been partly funded by the Spanish State Research Agency (AEI) Projects No.ESP2017-87676-C5-1-R and No. MDM-2017-0737 Unidad de Excelencia "Mar\'ia de Maeztu"- Centro de Astrobiolog\'ia (INTA-CSIC).

\end{acknowledgements}

\bibliographystyle{aa} 
\bibliography{occasoiv}

\appendix

\section{Details on data reduction}\label{apex:data_reduction}

\subsection{Sky emission and telluric absorption correction}

\begin{figure}
	\includegraphics[width=\columnwidth]{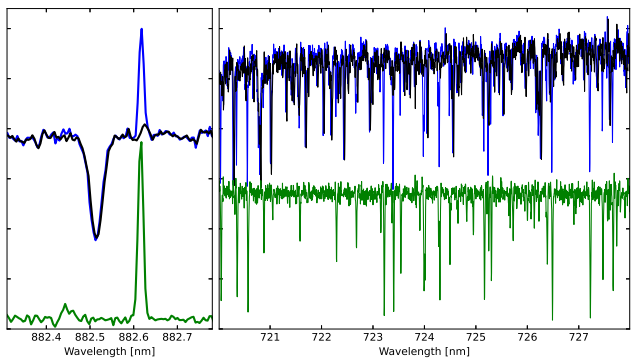}
    \caption{Example of the sky (left) and telluric subtraction (right). In both panels, blue and black lines are the spectrum before and after subtraction, respectively. Green lines are the subtracted spectra: sky emission and telluric absorption lines in left and right panels, respectively.}
    \label{fig:sky_tel_subtraction}
\end{figure}

As explained in Paper\,I, we acquire a blank field sky exposure in each run to subtract the sky emission. Moreover, we acquire several exposures of  bright, hot, and rapidly rotating stars to remove the telluric absorption features, such as bands of O$_{\rm 2}$ and H$_{\rm 2}$O.

The sky emission, both continuum and lines, is removed following the procedure described by \citet{carrera_king1}. For each order, the sky and object spectra are separated into two components: continuum and lines. The sky-line component is cross-correlated with the object-line one to put both in the same wavelength scale. This also provides an additional check of the wavelength calibration. The obtained offset are insignificant, to the order of 0.001 pixels, confirming no issues in the wavelength calibrations. Because object and sky exposures were not acquired in the same conditions, the sky- and object-line components are compared to search for the scale factor that minimizes the sky-line residuals over the whole spectral region covered by each order. In practice, this optimum scaling factor is the value that minimizes the sum of the absolute differences between the object-line and the sky-line multiplied by the scale factor, known as L1 norm. The object-continuum is added back to the sky-subtracted object-line spectrum. Finally, the sky continuum is subtracted, assuming that the scale factor is the same as for the sky-line component. As our goal is not to obtain flux calibrated spectra, and to avoid adding noise to the spectra, this procedure is only applied to those orders that contain significant sky emission lines, that is 3-$\sigma$ above the continuum level. Left panel of Fig.~\ref{fig:sky_tel_subtraction} shows an example of the sky subtraction performance. 

All the exposures of a telluric star are averaged to improve the signal-to-noise ratio. The continuum and stellar lines are removed for each order in the average telluric spectra in order to obtain only the telluric-line contribution. In the same way, the continuum of the object spectrum is removed in order to obtain the object-line contribution. As in the case of sky emission, the telluric- and object-line components are compared to search for the scale factor that minimizes the telluric line residuals over the whole spectral region covered by each order. After applying this scale factor, the telluric-line spectrum is subtracted from the object-line one before adding back the object-continuum. An example of the telluric subtraction is shown in the right panel of Fig.~\ref{fig:sky_tel_subtraction}.  

Finally, with each spectrum still separated in orders, the heliocentric correction is applied.

\subsection{Combination, normalization and merge}

\begin{figure}
	\includegraphics[width=\columnwidth]{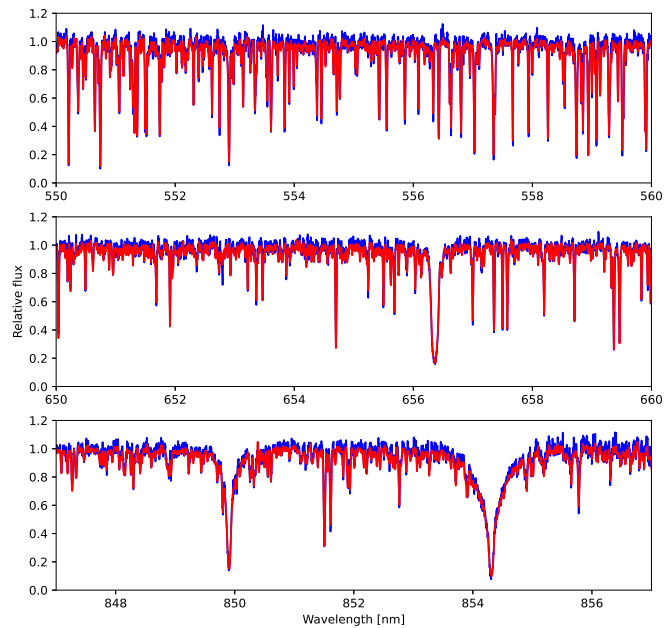}
    \caption{Comparison between the final spectra obtained with the old in blue (see Paper\,I for details), and new, in red, procedure in three different wavelength windows for a typical star acquired with HERMES at Mercator telescope.}
    \label{fig:red_comb}
\end{figure}

At least three exposures of each individual target are acquired with the goal of removing cosmic rays contamination. The procedure followed to combine them is the following. (i) The exposure with the highest S/N is chosen as reference. With the individual spectra still separated by orders, each order of each individual exposure is cross-correlated with the same order of the reference exposure to determine initial shifts. (ii) After applying these shifts for each order, the individual exposures are averaged using a $\kappa$-sigma clipping rejection and weighting by the individual S/N of each exposure. (iii) Again, for each order, the individual spectra are cross-matched with the averaged one in order to refine the shifts between them. Steps (ii) and (iii) are repeated until no significant shifts are found. This procedure increases the S/N of the averaged spectra, but also allows detecting radial velocity variability if the exposures are properly distributed (see next section).

The averaged spectra of each order is normalized by fitting the upper envelope with a low degree polynomial to remove the effect of the instrument response in the shape of the spectra. It is difficult to fit several very crowded orders and orders that contain strong lines. To address this issue we take advantage of the fact that the shape of each order, which is mainly determined by the instrument sensitivity, is related to that of the adjacent orders. Therefore, the continuum of the problematic orders is obtained by interpolation of the continuum of the rest of the orders. This procedure is iteratively repeated until no significant relative variations with respect to the previous iteration are found. Finally, the individual orders are merged in a single 1D spectrum. The flux of the overlapping regions covered by two orders are obtained by averaging both. Figure~\ref{fig:red_comb} shows an example of a final obtained spectrum, in red, and the comparison with the result of the previous procedure, in blue (see Paper\,I for details). 

\section{Notes on comparison with external surveys}\label{apex:comp_svry}

\subsubsection{Gaia DR2}

The \gaia DR2 \citep{gaia2018dr2} provides radial velocities for more than 7 million stars \citep[][]{gaiadr2rv} determined with Radial Velocity Spectrometer \citep[RVS,][]{gaiadr2rvs}. Its wavelength coverage is from 845 to 872\,nm with a spectral resolution of 11\,500. The precision of the obtained radial velocities depends on several features such as the magnitude of the targets but also on their temperature, with values between $\sim$1 to 4\kms\citep[][]{gaiadr2rv}. In total, there are 238 OCCASO stars with radial velocities in \gaia DR2. The differences of the radial velocities show a Gaussian distribution (panel a of Fig.~\ref{fig:comp_surveys1}) with a median of -0.04\kms. The relatively large width of the distribution has a MAD of 0.34\kms~and a standard deviation of 0.56\kms. Therefore, there is a good agreement between OCCASO and \gaia DR2 radial velocities in spite of the larger uncertainties involved in the \gaia measurements.

\subsubsection{\citet{mermilliod2008_RVOC}}

\citet{mermilliod2008_RVOC} determined mean radial velocities for 1\,309 stars belonging to 166 open clusters. The sample covering both hemispheres were obtained with two twin instruments, the photoelectric scanner CORAVEL \citep{coravel} installed at the Swiss 1\,m telescope at the Haute-Provence Observatory (France) and the Danish 1.54\,m telescope at La Silla Observatory (Chile). On average, the determined radial velocities have an internal dispersion of $\sim$0.5\kms. We have found 78 stars in common with OCCASO. The difference of the radial velocities obtained in both studies shows a well-defined peak (panel b of Fig.~\ref{fig:comp_surveys1}) with a median value of 0.29\kms~with a MAD of 0.12\kms~and a standard deviation of 0.22\kms. We performed the same comparison in Paper\,I finding similar results but from 40 stars. Therefore, we can conclude that the OCCASO radial velocities are in good agreement with \citet{mermilliod2008_RVOC} within the involved uncertainties. 

\subsubsection{APOGEE DR16}

The APOGEE \citep[][]{apogee} is one of the surveys performed in the framework of the third and fourth phases of Sloan Digital Sky Survey \citep[SDSS,][]{sdss_3,sdss_4}. APOGEE has obtained in both hemispheres R$\sim$22\,500 spectra in the infrared $H$-band, 1.5-1.7\,$\mu$m. The radial velocities determined by APOGEE have an uncertainty of $\sim$0.1\kms~ \citep{apogee_nidever}. We cross-match OCCASO with the latest APOGEE data release, sixteenth\citep [hereafter referred as APOGEE DR16,][]{sdss_dr16,apogeedr16}. We used a search radius of 5\arcsec\footnote{This is the angular distance that provides the best cross-match between Gaia DR2, the main coordinates source for our targets, and 2MASS, the APOGEE coordinates source \citep[see][for details]{gdr2_xmatch}.}. We have 50 stars in common with APOGEE. The difference of radial velocities shows a distribution peaked at -0.23\kms~with a median absolute deviation of 0.13\kms~(panel c of Fig.~\ref{fig:comp_surveys1}). Differences in this range were already reported by \citet{apogee_nidever}. Therefore, it is due to the existence of small systematics in the APOGEE radial velocity determinations.

\subsubsection{WIYN Open Cluster Study (WOCS)}

The WOCS (WIYN Open Cluster Study) is systematically  performing a comprehensive photometric, astrometric, and spectroscopic study of selected open clusters using the WIYN (Wisconsin, Indiana, Yale NOAO) 3.5\,m telescope at Kitt Peak Observatory (Arizona, USA). In particular, four of the clusters studied by WOCS are in common with OCCASO: NGC\,188 \citep{geller2008ngc188}, NGC\,2682 \citep{geller2015m67}, NGC\,6791 \citep{tofflemire2014ngc6791}, and NGC\,7789 \citep{nine2020ngc7789}. Although not always the same instruments and instrumental configurations have been used in all these works, their radial velocities have average uncertainties below 0.5\kms. We have found 33 stars in common with WOCS. The distribution of the radial velocities differences shows a well-defined peak (panel d of Fig.~\ref{fig:comp_surveys1}) with a median of 0.19\kms, MAD of 0.15\kms, and a standard deviation of 0.22\kms. Therefore, we can conclude that there is a good agreement between OCCASO and WOCS radial velocities within the uncertainties involved in each of the studies.

\subsubsection{LAMOST DR5}

The LAMOST (Large Sky Area Multi-Object Fiber Spectroscopic Telescope) is performing a Galactic Survey, also called as LEGUE (LAMOST Experiment for Galactic Understanding and Exploration survey). The LAMOST system is a 4\,m telescope which feeds a highly-multiplexed spectrograph in which 4000 fibers can be distributed on a 5\degr~field of view. LAMOST is sampling Galactic stars with two resolutions. The lowest one, R$\sim$1\,500, provides a full spectral coverage between  369 and 910\,nm. The highest resolution, R$\sim$7\,500, provides spectra in two bands: blue, 495 to 535\,nm; and red, 630 to 680\,nm. For the sixth data release, DR 6, the radial velocity uncertainties in the lowest resolution mode are, on average, of about 5\kms~while in the highest one they are, on average, of 1.5\kms. There are 32 stars in common between OCCASO and LAMOST, but there is no significant agreement as shown in panel e of Fig.~\ref{fig:comp_surveys1}  probably due to the LAMOST large uncertainties. The median of the radial velocities differences is 5.5\kms~with a MAD of 2.5\kms~and a standard deviation of 3.8\kms.

\subsubsection{\gaia RVS standards}

\citet{soubiran_rvstd} compiled a catalogue of 4\,813 stars used as reference for the \gaia RVS. In total, they have compiled $\sim$71\,000 radial velocities measurements from five high resolution spectrographs. The resulting radial velocities have a typical time baseline of 6\,a with a MAD of 15\,m\,s$^{-1}$. There are 28 stars in common with OCCASO. The differences of the radial velocities show a clear narrow peak centred at 0.22\kms~with a MAD of 0.03\kms~and a standard deviation of 0.06\kms~(panel f of Fig.~\ref{fig:comp_surveys1}). There may be a systematic between the radial velocity determination of both samples. Therefore, there is a good agreement between both samples, but with a systematic error of $\sim$0.2\kms.

\subsubsection{GALAH DR3}

GALAH is a large observing program using the HERMES \citep[High Efficiency and Resolution Multi-Element Spectrograph,][]{hermes_aat} instrument with the Anglo-Australian Telescope of the Australian Astronomical Observatory. HERMES provides simultaneous medium resolution, R$\sim$28\,000, spectra for 400 objects in four wavelength bands: blue (472-490\,nm), green (565-587\,nm), red (648-674\,nm), and infrared (759-789\,nm). The third data release of GALAH includes also the radial velocities determined by the SME (Spectroscopic Made Easy) pipeline \citep[][]{GALAHDR3}. The typical uncertainty of the GALAH DR3 radial velocities is $\sim$0.1\kms. There are 25 stars in common between OCCASO and GALAH EDR3 in spite of GALAH mainly sampling the Southern Hemisphere. The difference of the radial velocities determined by both projects does not show a peaked distribution (panel g of Fig.~\ref{fig:comp_surveys1}). The median of the differences is -1.09\kms~but with a median absolute distribution 0.80\kms~and a standard deviation of 1.40\kms. These large values denote the poor agreement between the OCCASO radial velocities and the values provided by GALAH DR3

\subsubsection{Gaia-ESO Survey (GES) DR4}

GES \citep{ges_gilmore} is a spectroscopic survey carried out with FLAMES \citep[Fibre Large Array Multi Element Spectrograph][]{flames} on one of the VLT units (Very Large Telescopes). FLAMES is a multi-fibre instrument which feeds two different spectrographs: UVES and GIRAFFE. UVES \citep[Ultraviolet and Visual Echelle Spectrograph,][]{uves}
is a high-resolution, R$\sim$47\,000, optical spectrographer covering a wavelength range from 480 to 700\,nm which can be feed with up to 8 fibres. The other 132 fibres go to GIRAFFE\footnote{Its name is related to the first concept of this instrument.}, a medium-resolution, R$\sim$20\,000 spectrographer which is able to cover the whole visible wavelength range using different gratings. However, different combinations of them have been used for GES-GIRAFFE and therefore, there is no specific configuration for the entire survey. This implies that not all target have been sampled in the same way as consequence the radial velocities uncertainties are between 0.15 and 0.37\kms~\citep[e.g.][]{lardo2015gesgc}. There are only 20 OCCASO stars in common, owing to GES observed from the Southern Hemisphere. The difference of the radial velocities shows a bimodal distribution peaked at -0.37 and 0.53\kms~and with a $\sigma$ of 0.19 and 0.15\kms, respectively (panel h of Fig.~\ref{fig:comp_surveys1}). None of the peaks is related to a particular instrumental configuration of GES, which could explain the obtained distribution. Therefore, GES DR3 radial velocities show a larger scatter in comparison with the OCCASO ones.

\subsubsection{RAVE DR6}

The RAVE \citep[Radial Velocity Experiment,][]{rave} is a magnitude-limited (9$<I<$12\,mag) spectroscopic survey of Galactic stars randomly selected in the Southern Hemisphere. It obtains low-resolution, R$\sim$7\,500, spectra covering the infrared \ion{Ca}{ii} triplet region, 841 to 879\,nm, using the 6df multi-object spectrograph \citep[6\degr~diameter field of view,][]{6df_instrument} installed at the UK Schmidt Telescope (UKST) in Australia. There are 16 stars in common between OCCASO and the sixth RAVE data release. The distribution of the radial velocity differences (panel i of Fig.~\ref{fig:comp_surveys1}) does not show a clear peaked distribution, its median is -0.22\kms~with a MAD of 0.85\kms~and a standard deviation of 1.00\kms.

\subsubsection{The Geneva-Copenhagen survey}

\citet{nordstrom2004} presented the Geneva-Copenhagen survey of the Solar neighbourhood in which radial velocities were determined for $\sim$14\,000 F- and K-type stars. The derived radial velocities have, on average, an uncertainty of $\sim$0.3\kms. OCCASO  has 13 stars in common with this survey. The median of the radial velocities differences is of 0.64\kms~with a MAD of 0.30\kms~and a standard deviation of 0.45\kms~(panel j of Fig.~\ref{fig:comp_surveys1}).  

\subsubsection{SEGUE}

SEGUE \citep[Sloan Extension for Galactic Understanding and Exploration,][]{segue} obtained low resolution, R$\sim$1\,800 spectra with a wavelength coverage from 390 to 900\,nm for Galactic stars in the range 14$<g<$20\,mag in the framework of the SDSS. The determined radial velocities have an uncertainty which ranges from $\sim$4\kms~at $g<$18\,mag to $\sim$15\kms~at $g\sim$20\,mag. In spite of SEGUE have sampled some stars in several open clusters \citep[e.g.][]{smolonski2011segueoc,carrera2012ngc6791,carrera2013}, we have found only 5 stars of the cluster NGC\,6791 in common with OCCASO. We found a median difference of 0.55\,\kms~with a MAD of 0.22\kms~and a standard deviation of 0.43\kms~which is a good agreement taken into account the uncertainties involved in the SEGUE radial velocities.

\section{Notes on individual clusters}\label{apex:indvnotes}

We began this project before the \gaia results, so we have done now an updated study taking into account the new data. In the following, when we refer to the astrometric membership probabilities, $p$, we refer to the values computed by \citet{cantatgaudin2018} and \citet{cantatgaudin2020} from \gaia DR2 proper motions and parallaxes. We refer the reader to these papers for details. Moreover, except if otherwise specified, all the \gaia identifications refer to EDR3.

\begin{figure*}[!]
\centering
\includegraphics[width=\textwidth]{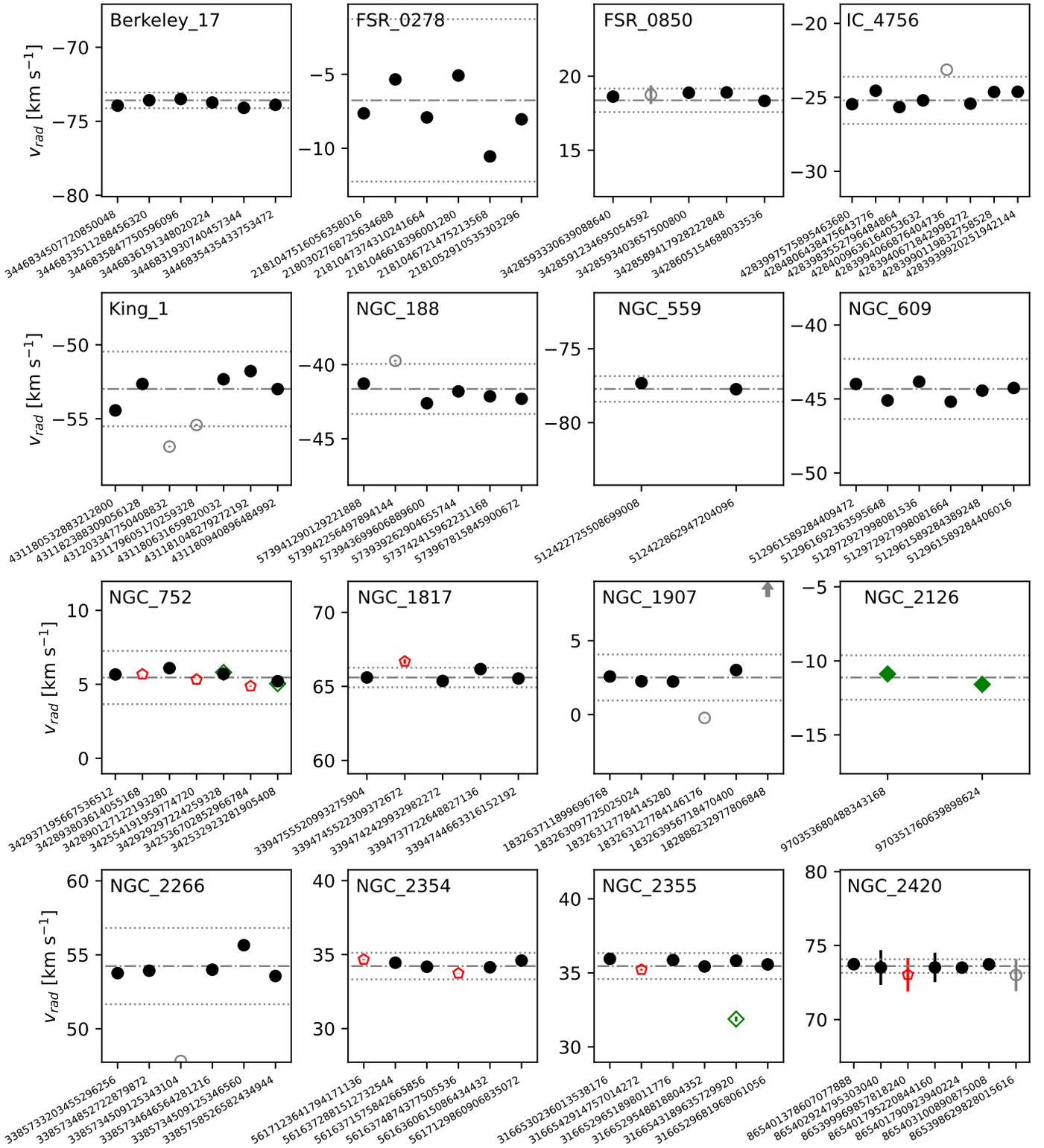}
\caption{As Fig.~\ref{fig:rv_membership} for the remaining clusters.}
\label{fig:apex_rv_mermbership}
\end{figure*}

\begin{figure*}[!]
\addtocounter{figure}{-1}
\centering
\includegraphics[width=\textwidth]{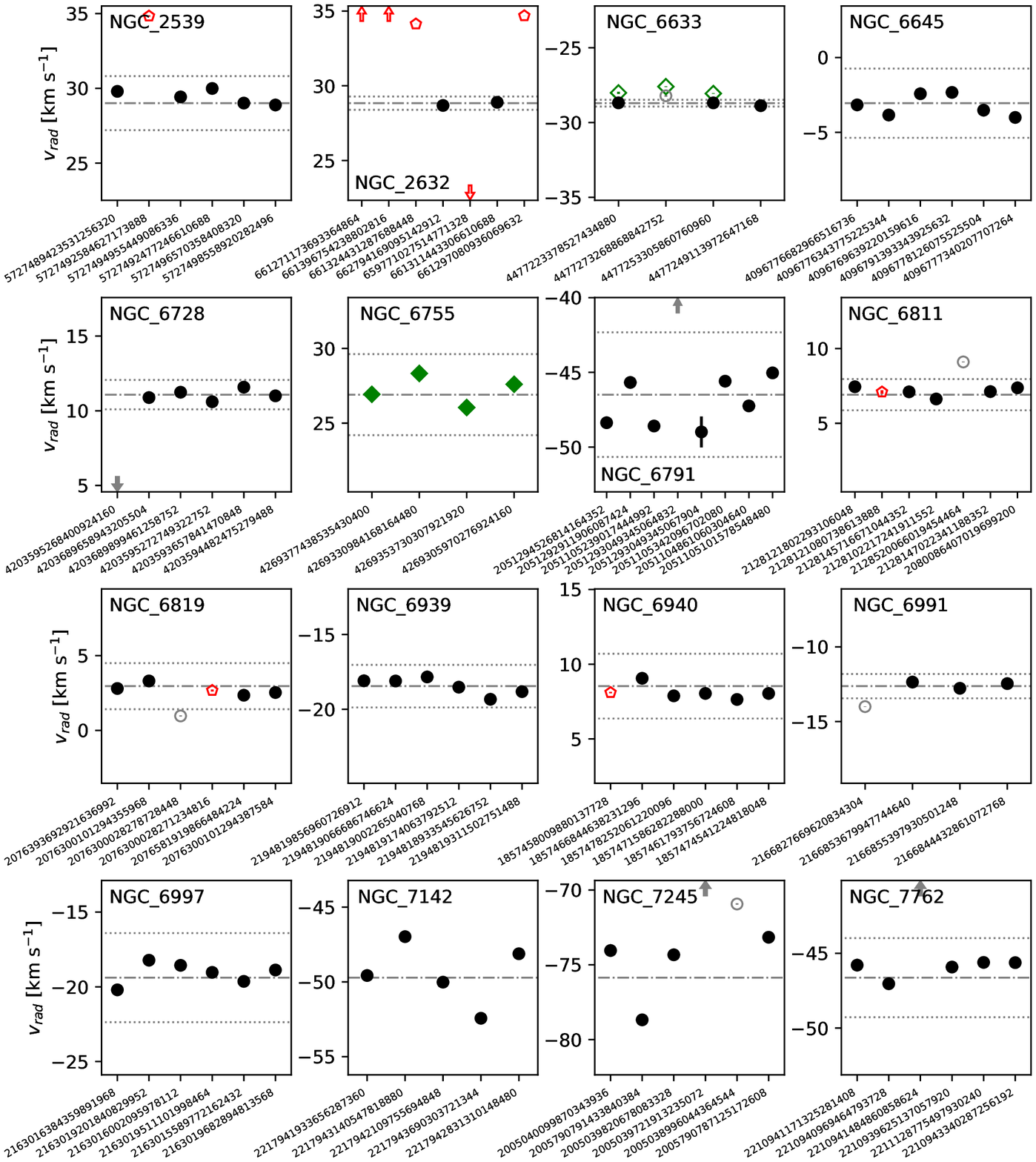}
\caption{Cont.}
\end{figure*}

\begin{figure*}[!]
\addtocounter{figure}{-1}
\centering
\includegraphics[width=\textwidth]{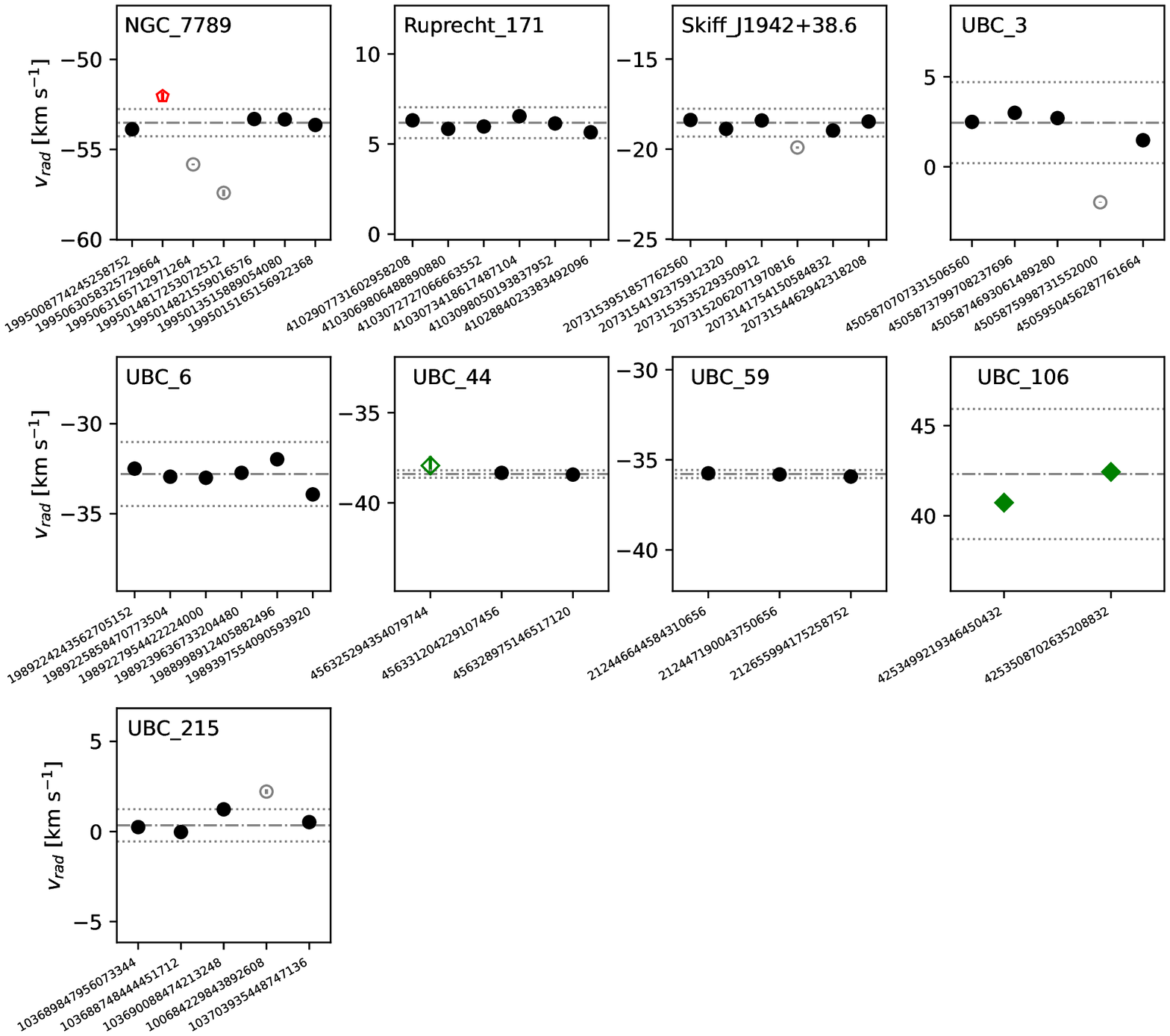}
\caption{Cont.}
\end{figure*}

\begin{figure}
\centering
\includegraphics[width=\columnwidth]{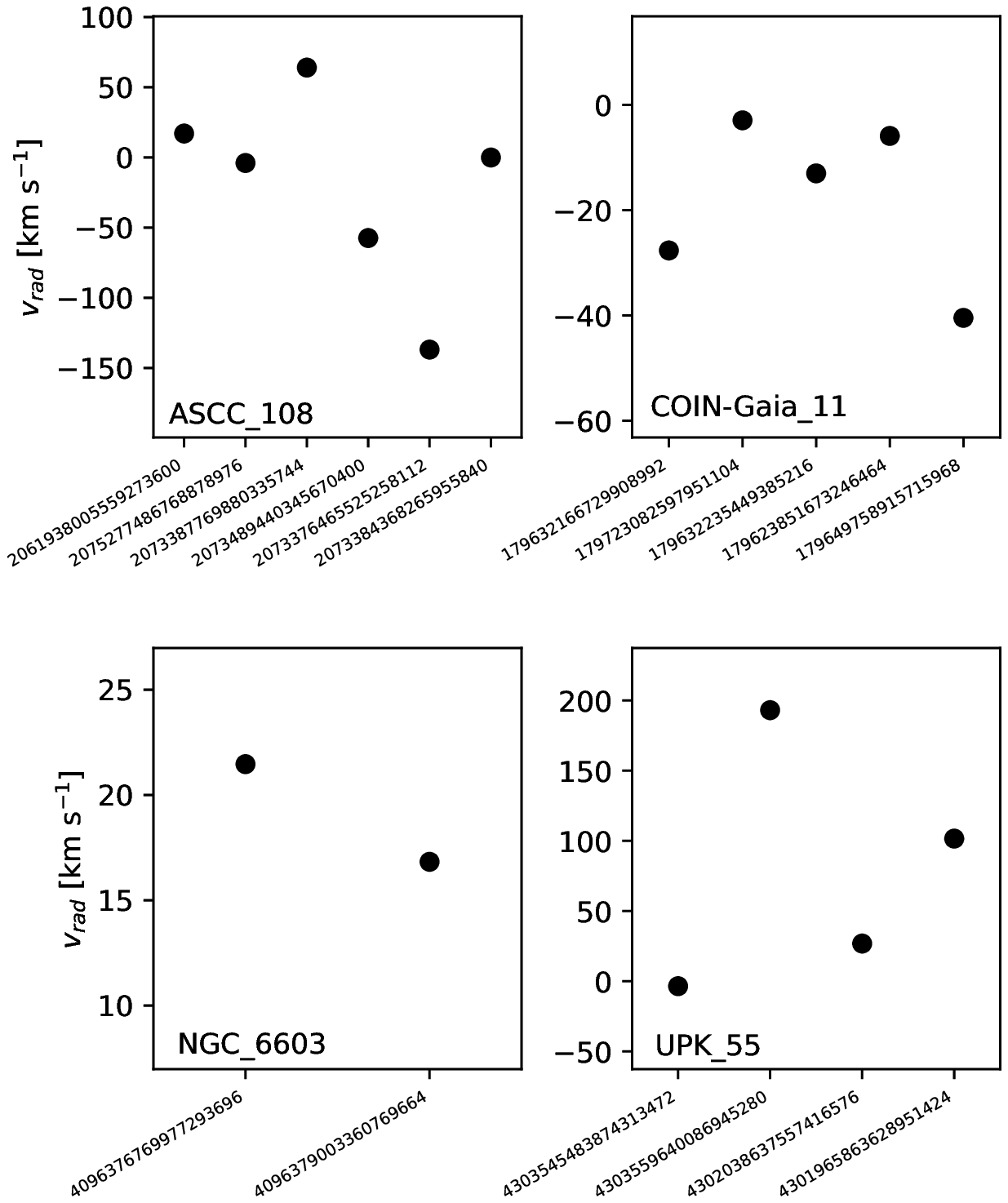}
\caption{As Fig.~\ref{fig:rv_membership} but for the clusters for which we are not sure of targeting real cluster members. Note that the scales of y-axis are much longer than in Fig.~\ref{fig:rv_membership} and change from panel to panel.}
\label{fig:rv_nomembership}%
\end{figure}

\subsection{ASCC\,108}

We have targeted six stars in the field of view of this cluster. Four of them have $p\geq$0.7, but the other two have a negligible probability of being cluster member. In any case, all the studied stars have very different velocities among them, even those with a high membership probability. \citet{soubiran2018gaiadr2_rv} reported a radial velocity for this cluster of -6.3$\pm$2.1\,km\,s$^{-1}$ but from a single star which have not been targeted by OCCASO. Therefore, we are not able to ensure that OCCASO has target real cluster single members. For this reason we do not provide a radial velocity for this cluster.

\subsection{Alessi~1}

For the moment, we have observed a single star in this cluster, but with $p$=1. Although this star has been observed with CAH2, which implies larger uncertainties, the derived radial velocity is similar to the value obtained by \gaia RVS. Moreover, this value, is in agreement with the average velocity obtained for this cluster by \citet{soubiran2018gaiadr2_rv} from 13 stars of -3.73$\pm$1.42\,km\,s$^{-1}$. Therefore, we assign to this cluster the radial velocity of this star of -4.67$\pm$0.03\,km\,s$^{-1}$ taken into account that only one star has been observed and the real uncertainty of the OC may be of $\sim$0.5\,km\,s$^{-1}$.

\subsection{Berkeley~17}

All the six stars observed in Berkeley~17 have $p$=1. Three of the stars have slightly different radial velocities than the others: W117, W120 and W130. These three have $v_{scatter}$ values larger than the average of the sample, which could be a hint of being spectroscopic binaries. Although these stars are among the faintest objects in our sample and therefore, this may increase the uncertainties in their radial velocity determinations. Excluding these three stars the average cluster radial velocity is -73.59$\pm$0.03\,km\,s$^{-1}$ while using the three stars the value is: -73.79$\pm$0.09\,km\,s$^{-1}$.

\subsection{COIN-Gaia~11}

Five stars have been observed in the field of view of COIN-Gaia~11, one of the new clusters reported by \citet{cantat_gaudin2019coin} from \gaia DR2. Three of the observed stars were selected because they are located near the expected position of the red-clump, although they have a very low membership probability, $p\leq$0.2. The other two stars have $p$=0.6 and 0.8, respectively \citet{cantatgaudin2018}. The observed stars have very different radial velocities among them. Therefore, we cannot ensure that we are observing real cluster single members.

\subsection{FSR~278}

The six stars observed in the field of view of FSR~278 have $p\geq$0.7, but they show slightly different radial velocities among them. To our knowledge, there are no previous radial velocity determinations for this cluster in the literature. For this reason we provided the obtained value in spite of its large scatter ($-$6.76$\pm$1.83\,km\,s$^{-1}$). The observed scatter could be explained if our sample will include one or more spectroscopic binaries, a statement that could not be checked with the data already on hand.  

\subsection{FSR~850}

The five sampled in FSR\,850 have probabilities in the range 0.5$\leq p\geq$0.9. All of them have similar radial velocities within the uncertainties. There is one star, 3428591234695054592, shows a significant scatter in the radial velocities of their individual exposures, $v_{scatter}$=0.63\kms, which is an indication of being an spectroscopic binary.

\subsection{IC~4756}

The eight observed stars in this cluster have $p$=1 except 4283940671842998272 (W0081) that has proper motions and parallaxes incompatible with the cluster  \citep[][]{cantatgaudin2018}. This star was selected before \gaia based on its position in the colour-magnitude diagram (Fig.~\ref{fig:dcms}). However, in Paper\,I we already reported that it has a radial velocity different than the other cluster's stars.

\subsection{King~1}

Two of the seven stars observed in King~1 have incompatible proper motions and parallaxes from EDR3 with the average values of the cluster: 431203347750408832 (W0405) and 431179605170259328 (W0971), as well as slightly different radial velocities. These stars were selected before the Gaia data was available and are clearly non-members. The other observed stars have $p$=1 and similar radial velocities, although with a dispersion larger than the typical values observed in other clusters.

\subsection{Melotte\,72}

For the moment, we have observed a single star in this cluster with $p$=1. For this star we obtained a radial velocity of 70.70$\pm$0.11\kms. This value is compatible with the average radial velocity provided by \citet{soubiran2018gaiadr2_rv} for this cluster of 71.50$\pm$\.07\kms~from 8 stars.

\subsection{Melotte~111}

We have observed 12 stars along the main sequence of the Coma Berenices Cluster (Melotte~111 or Collinder~256). There are two giant stars with a high membership probability in this cluster reported by \citet{cantatgaudin2020}: 4008342726516877312 and 3961912034103154944. The first one is a spectroscopic binary \citep[][]{mermilliod2009} and the former a well-known rotationally variable star \citep[][]{massaroti2008}. For these reasons we have not targeted them. Two of the observed stars: 4002275552635131136 (W058) and 4008342623437661568 (W092); have a negligible membership probabilities from \citet{cantatgaudin2020}. Their  \gaia EDR3 proper motions are incompatible with the average values of the cluster, although they have significantly larger uncertainties in comparison with the bulk of the cluster. Both objects have radial velocities compatible with the average value for the cluster, although the value for the first stars is slightly larger but still compatible of being a cluster member within  uncertainties. For this reason, we consider these two stars as cluster members from their radial velocity. There are two stars, with $p$=1, with radial velocities statistically different from the rest of the objects sampled in this system: 4002565308308607616 (W036) and 4002550293102977152 (W049). The \gaia~DR2 radial velocities are also different from the rest of the cluster members \citep{soubiran2018gaiadr2_rv}. The first object has a significantly larger rotational velocity, $v_{sini}$ according to \citet{mermilliod2009} than the other objects sampled in this cluster. To our knowledge, there are no recent measurements of the rotational velocity of the other star. The current version of our radial velocity determination pipeline does not include determination of the rotational velocity. However, these two stars have the largest  $v_{err}$ values in our sample. As $v_{err}$ is the width of the correlation peak, larger values could imply that the lines are wider than the template ones, a clear sign of rotation. For the moment, we discard these two objects as cluster members from their radial velocity.

\subsection{NGC~188}

The six stars targeted in NGC~188 have $p$=1. All of them have compatible radial velocities within the uncertainties although one of them, 573942256497894144 (W2051), has a radial velocity slightly different from the other but still within 3~$\sigma$. We do not find in the literature that this could be a spectroscopic binary or a high rotation star. This star has also been observed by APOGEE DR16 which reported a velocity similar to the one obtained here: $v_{APO,DR16}$=-40.06$\pm$0.039\,km\,s$^{-1}$.

\subsection{NGC~559}

For the moment we have sampled only two stars in this cluster, all of them we $p$=1 and both with compatible radial velocities. We have used both in our analysis. 

\subsection{NGC~609}

The six stars observed in NGC\,609 have $p\geq$0.9. All of them have compatible radial velocities within the uncertainties.

\subsection{NGC\,752}

The seven stars observed have $p$=1. \citet{mermilliod2008_RVOC} reported three of them as spectroscopic binaries: 342554191959774720 (W0001), 342536702852966784 (W0027), and 342893803614055168 (W0295). They show radial velocities compatible with the average values of the cluster. We exclude them from our analysis.

\subsection{NGC~1817}

Four of the five stars observed in NGC~1817 have $p\geq$0.9 \citep{cantatgaudin2018}. The exception is the star 3394745522309372672 (W0022) with $p$=0.5 which has been reported as spectroscopic binary by \citet{mermilliod2008_RVOC} although it has a radial velocity compatible with the other stars in the cluster.

\subsection{NGC~1907}

One of the five stars observed in NGC~1907, 182888232977806848 (W2087), has \gaia EDR3 proper motions and parallax incompatible with the average value of the cluster and also a very different radial velocity. It was selected because of its position in the colour-magnitude diagram. The other four objects have $p\geq$1 except for star 183263127784146176 (W0133) with $p$=0.8. This star has a radial velocity slightly different that the others and outside the 3~$\sigma$ range. A slightly different radial velocity for this star was already reported in Paper~I. We consider this star as cluster non-member, although it could be a spectroscopic binary.

\subsection{NGC~2099}

The 12 observed stars in this cluster have $p\geq$0.9 and all of them have compatible radial velocities within the uncertainties. However, the star 3451180838527642496 observed with CAH2 has not been used to obtain the average radial velocity and dispersion of the cluster.

\subsection{NGC~2126}

The two objects observed in NGC~2126, which have $p$=1, have similar radial velocities within the uncertainties in spite of being observed with CAH2. We have obtained the average radial velocity from them, although the obtained value may be taken with care.

\subsection{NGC~2266}

Although the six observed stars in this cluster have $p$=1 one of them, 3385734509125343104, has a radial velocity slightly different than the others. Therefore, we consider this star as non-member from radial velocity.

\subsection{NGC~2354}

The six observed stars in NGC~2354 have $p$=1 although two of them, 5616374874377505536 (W125) and 5617123641794171136 (W183), are catalogued as spectroscopic binaries by \citet{mermilliod2008_RVOC}. We excluded them in our analysis, but they have radial velocities compatible with the rest of the cluster.

\subsection{NGC~2355}

In NGC~2355, the six targeted stars have $p\geq$0.8 although one of them, 3166542914757014272 (W536), has been reported as spectroscopic binary by \citet{mermilliod2008_RVOC}. We discard this object in our analysis, although it has a radial velocity compatible with the other studied stars. 

\subsection{NGC~2420}

The seven stars observed in NGC~2420 have $p\geq$0.9. One of them, 865399969857818240 (W091), has been reported as spectroscopic binary by \citet{mermilliod2008_RVOC}. Other stars in this cluster are affected by the problems in the wavelength calibration of the first observing run with NOT1 in April 2013 which is the source of their large error bars. Because of this problem one of these stars, 865398629828015616 (W236), is excluded because its radial velocity is outside the 3~$\sigma$. Although with its large uncertainty the derived velocity is compatible with the average value for the cluster, we excluded it so as not to artificially increase the internal velocity dispersion in the cluster. However, although we excluded it of our analysis, we keep it as a potential cluster member for future chemical analysis.

\subsection{NGC~2539}

The six stars observed in this cluster have $p$=1 although one of them, 5727492584627173888 (W233) may be a spectroscopic binary according to \citet{pourbaix2004} and its radial velocity is not compatible in our sample. 

\subsection{NGC\,2632}

We have observed 7 stars in NGC\,2632, commonly known as Praesepe, all of them with $p\geq$0.9. However, five of them have been reported as spectroscopic binaries by \citet{mermilliod2008_RVOC} and \citet{pourbaix2004}: 661297080936069632, 661271173693364864, 661396754238802816, 661324431287688448, and 659771027514771328, and their radial velocities are not compatible. The average radial velocity for this cluster has been derived from the remaining two stars, which have very similar values.

\subsection{NGC\,2682}

A total of 14 stars with $p\geq$0.7 have been observed in NGC\,2682, also known as M\,67. Three of them have been reported as spectroscopic binaries by different sources: 604904503934969856 (W224), 604917835513458688, and 604917354477128448 (W170) \citep{mathieu1990,mermilliod2008_RVOC,geller2015m67,geller2021_m67}. The star 604904503934969856 has a large $v_{scatter}$ which is a clear indication of being a binary as the large error bar in Fig.~\ref{fig:rv_membership} denotes. We have excluded these three stars of the analysis. There is another star 604917629355042176 (W141) which was reported as binary by \citet{mermilliod2008_RVOC} but not in more recent studies \citep[e.g.][]{geller2015m67,geller2021_m67}. Its radial velocity is in very good agreement with the other stars sampled in the cluster. Therefore, we fully consider this star in our analysis. There are four stars observed with CAH2 which have been excluded of the determination of the average radial velocities of the cluster, although they are considered as member since their velocities are compatible to that of the cluster within the uncertainties.

\subsection{NGC~6603}

For the moment we have observed only two stars in NGC~6603, 4096376769977293696 (W2033) and 4096379003360769664 (W2252) with $p$=0.8 and 0.7, respectively. Both stars have different radial velocities. For this cluster, \citet{soubiran2018gaiadr2_rv} reported an average velocity of 17.9$\pm$1.4\,km\,s$^{-1}$ from the radial velocities of 14 stars released in \gaia DR2. We have only one star, 4096379003360769664, in common with the \gaia sample which provided a radial velocity in spite of the larger uncertainties, 16.77$\pm$0.14\,km\,s$^{-1}$, very similar to the value found here, 16.83$\pm$0.01\,km\,s$^{-1}$. In any case, because the two sampled stars have statistically different radial velocities, we do not provide an average radial velocity for this cluster.

\subsection{NGC~6633}

The four stars observed in NGC~6633 have astrometric probabilities between 0.5 and 0.9. The four stars have similar radial velocities. However, because of the small error bars on one of the stars, 4477273268868842752 (W106), with the lowest membership probability, $p$=0.5, is rejected by the 3~$\sigma$ clipping. However, because of the very small error bars involved we do not discard that this object is a real NGC~6633 member.

\subsection{NGC~6645}

The six observed stars in NGC~6645 have $p\geq$0.8. All of them have compatible radial velocities within the uncertainties.

\subsection{NGC\,6705}

We have observed a total of 17 stars in this cluster. Except for 4252502851280454528 (W1256) which has $p$=0.4 the other stars have $p\geq$0.7. The star 4252499041749404288 (W1090) has been catalogued as spectroscopic binary by \citet{mermilliod2008_RVOC}, although the value found here is compatible with the other observed stars within the uncertainties. 

\subsection{NGC\,6728}

The six stars observed in NGC\,6728 have $p\geq$0.9. However, one of them, 4203595268400924160, has a different radial velocity in comparison with the other stars in this cluster. This star has been excluded in our analysis.

\subsection{NGC\,6755}

We have observed four stars with CAH2 in this cluster. They have $p\geq$0.8. All of them have compatible radial velocities within the uncertainties.

\subsection{NGC\,6791}

We have observed eight stars in NGC\,6791, all of them with $p$=1. They are among the faintest objects in our sample and therefore, they have lower S/N than the bulk of the objects. Moreover, one of the stars, 2051293049345067904 (W10806) is affected by the problems of the NOT~APR13 observing run as its large error bar denotes. The star 2051293049345064832 (W10435) has a radial velocity significantly different from the rest of the sampled objects, and we consider it as a cluster non-member. The radial velocity found here, -13.49$\pm$0.04\kms, is significantly different from the value found by \citet{tofflemire2014ngc6791} within WOCS of -43.1\kms\footnote{No uncertainty is provided by the authors for this star.} although they reported that these objects should be a rapid rotation star with $v_{sini} =$47.7\kms. Our analysis based on 4 individual exposures does not show sign of rotation with an $v_{err}$=0.0002\,km\,s$^{-1}$.

\subsection{NGC\,6811}

We have sampled seven highly probability, $p$=1, stars in the field of view of NGC\,6811. One of them, 2128520066019454464 (W359), has a radial velocity statistically different from the others. Another star, 2128121080738613888 (W0032), has been reported as spectroscopic binary by \citet{mermilliod2008_RVOC}. We found a radial velocity compatible with the other stars, however it shows a $v_{scatter}$ slightly larger than most of the objects in our sample, which may be a hint of binarity. We excluded both objects of our analysis.

\subsection{NGC\,6819}

The six observed stars have $p$=1 except star 2076393692921636992 (W386) which has $p$=0.8. Star 2076300028271234816 (W979) has been reported as spectroscopic binary by \citet{pourbaix2004} and \citet{milliman2014ngc6819} although its radial velocity is compatible with the average value for the cluster. The star 2076300028278728448 (W978) has a radial velocity outside the 3~$\sigma$ limit. APOGEE DR16 radial velocity for this star, 1.18$\pm$0.08\kms, is similar to the value found here, 0.97$\pm$0.04\kms, within the uncertainties. We discard this star in our analysis.

\subsection{NGC\,6939}

The six stars sampled in NGC\,6939, all of them with $p$=1, have also compatible radial velocities among them.

\subsection{NGC\,6940}

The six stars observed in this cluster have $p$=1. One of them, 1857458009880137728 (W105), has been marked as spectroscopic binary by \citet{mermilliod2008_RVOC}, and therefore excluded in our analysis, although it has a radial velocity compatible with the others.

\subsection{NGC\,6991}

Three of the four stars observed in NGC\,6991 have $p$=1. The other one, 2166827669620834304 (W034), has $p$=0.8. However, its radial velocity statistically differs from the average value of the cluster, and we exclude it from our analysis. This star was also reported as non-member in Paper~I.

\subsection{NGC\,6197}

The six stars sampled in this cluster has relatively low astrometric membership probabilities, 0.6$\leq p\leq$0.7. Considering all of them as cluster members, we find an average radial velocity of -19.39$\pm$1.00\kms. This value is in good agreement with the value derived by \citet{soubiran2018gaiadr2_rv} for this cluster, -19.6$\pm$0.6\kms, obtained from 8 stars, including  five of the six objects studied here.

\subsection{NGC\,7142}

One of the observed stars, 2217943140547818880 (W15) has $p$=0.5, while the remaining four have $p$=1. Two of them, 2217941933656287360 (W09) and 2217942109755694848 (W98), have relatively large values of $v_{scatter}$, 0.13 and 0.21\kms, respectively. This could be a hint of binariety. From the five observed stars we found an average velocity of -49.7$\pm$2.8\kms~in good agreement with the value reported by \citet{soubiran2018gaiadr2_rv}, -49.7$\pm$0.9\kms, from the \gaia DR2 radial velocities of 21 stars including the 5 objects studied here. 
There are significant differences, $\sim$3\kms, between the radial velocities derived here, and the values provided by \gaia DR2 for three of the stars: 2217943140547818880 (W15), 2217942109755694848 (W98), and 2217943690303721344.  

\subsection{NGC\,7245}

The selection of the targets in NGC\,7245 was performed before \gaia DR2. As a consequence, three of the six observed stars have very low membership probabilities: $p$=0.1 for 2005039721913235072 (W045), and $p$=0.2 for 2005038996044364544 (W055) and 2005039820678083328 (W205).
The first two objects have proper motions incompatible with the bulk of the cluster in \gaia EDR3. In fact, 2005039721913235072 has a significant different radial velocity. Therefore, we consider these two star as non-members. The other star, 2005039820678083328, has a radial velocity and \gaia EDR3 proper motions compatible with other objects in the cluster. The star 2005790791433840384 (W178) has a $p$=1, but its radial velocity is different from the other three stars, but we cannot discard it from the data already on hand.

\subsection{NGC\,7762}

Five of the six observed stars in this cluster have $p$=1 and compatible radial velocities. The remaining object, 2210941484860858624 (W084), is clearly a non-member since it has incompatible proper motions, parallax and radial velocity. 

\subsection{NGC\,7789}

Six of the seven stars observed in NGC\,7789 have $p$=1. The exception is the star 1995015165156922368 (W10915) with $p$=0.2. One of the stars, 1995063058325729664 (W7176), shows a large $v_{scatter}$ value, 0.22\kms, which could be a sign of being a spectroscopic binary. Two of the high probability stars have statistically discrepant radial velocities: 1995063165712971264 (W7714) and 1995014817253072512 (W8260). On the contrary, the star with the lowest priority has a radial velocity compatible with the average value of the cluster. We consider this star as member, and we excluded the two stars with discrepant radial velocities and the potential spectroscopic binary.

\subsection{Ruprecht\,171}

The six stars observed in Ruprecht\,171 have $p$=1 and radial velocities compatible within the uncertainties.

\subsection{Skiff\,J1942+38.6}

The six stars sampled in this cluster have $p$=1. Statistically, all of them have similar radial velocities within the uncertainties with an average value of -18.61$\pm$0.48\kms. However, there is one star, 2073152062071970816, which radial velocity is near the 3~$\sigma$ limit. If we exclude this star, the average cluster radial velocity is -18.53$\pm$0.25\kms.

\subsection{UBC\,3}

The paper where this cluster was firstly reported, \citet{castroginard2018}, provided astrometric membership probabilities, both with $p$=1, for only two stars at the red clump position: 4505873799708237696 and 4505874693061489280. We added three additional targets by selecting stars with parallaxes and proper motions compatible with those of the cluster within the uncertainties. Two of these additional targets have compatible radial velocities with the others: 4505870707331506560 and 4505950456287761664. On the contrary, the star 4505875998731552000 has a discrepant radial velocity and it is excluded from our analysis.

\subsection{UBC\,6}

As in the case of UBC\,3, astrometric membership probabilities were available only for two stars, 1989225858470773504 and 1989227954422224000, in UBC\,6 at the moment of performing the target selection. We selected four additional targets with compatible parallaxes and proper motions. All the additional targets have compatible radial velocities among them and with the high astrometric membership probabilities objects.

\subsection{UBC\,44}

The three stars observed in this cluster have $p$=1. All of them have similar radial velocities even if one of them, 456325294354079744, has been observed with CAH2 and, therefore, has larger uncertainties. We excluded this object to obtain the average radial velocity of the cluster.

\subsection{UBC\,59}

The three stars sampled in UBC\,59 have $p$=1 and very similar radial velocities.

\subsection{UBC\,106}

The two stars observed for the moment in UBC\,106 have $p$=1 and similar radial velocities within the uncertainties taking into account that they have been observed with CAH2. This cluster has been studied recently by \citet{negueruela_valparaiso1}, who named it as Valparaiso\,1. Their radial velocities determinations are in agreement with the values found here, and also with \gaia DR2, within the uncertainties.

\subsection{UBC\,215}

The five stars in UBC\,215 have $p$=1. One of them, 3100684229843892608, has a radial velocity outside the 3~$\sigma$ range and it is excluded. However, this star and 3103690088474213248 have large $v_{scatter}$ values, 0.13 and 0.27\kms, respectively. This is a sign that these two stars may be spectroscopic binaries.

\subsection{UPK\,55}

The four stars observed in this cluster have 0.7$\leq p\leq$0.9. However, they have very different radial velocities among them, and therefore we cannot ensure that any of them is a real cluster member.

\section{Evolution of $Z_{max}$ and eccentricity as a function of age for axisymmetric and non-axisymmetric potentials}\label{apex:zmax_e}

\begin{figure}
\centering
\includegraphics[width=\columnwidth]{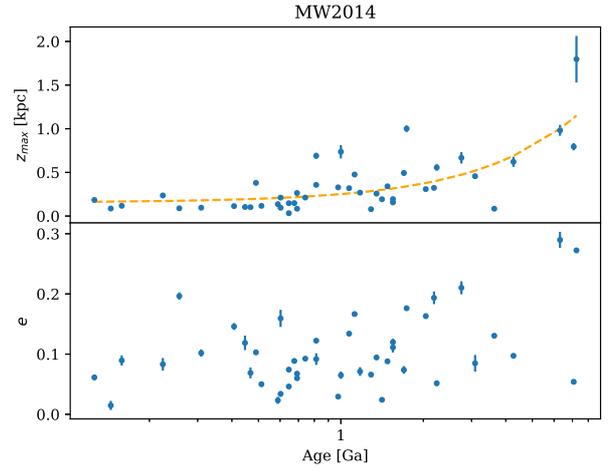}
\caption{As Fig.~\ref{fig:zmax_e-age} only with the axisymmetric potential.}
\label{fig:zmax_e-age_bar0}
\end{figure}

\begin{figure}
\centering
\includegraphics[width=\columnwidth]{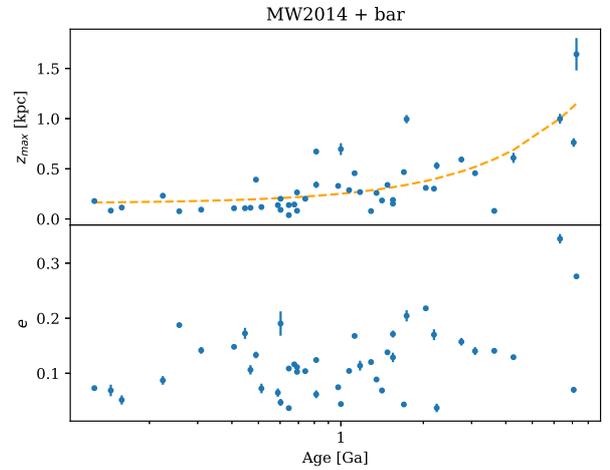}
\caption{As Fig.~\ref{fig:zmax_e-age_bar0} but adding the potential of the bar.}
\label{fig:zmax_e-age_bar1}
\end{figure}

\begin{figure}
\centering
\includegraphics[width=\columnwidth]{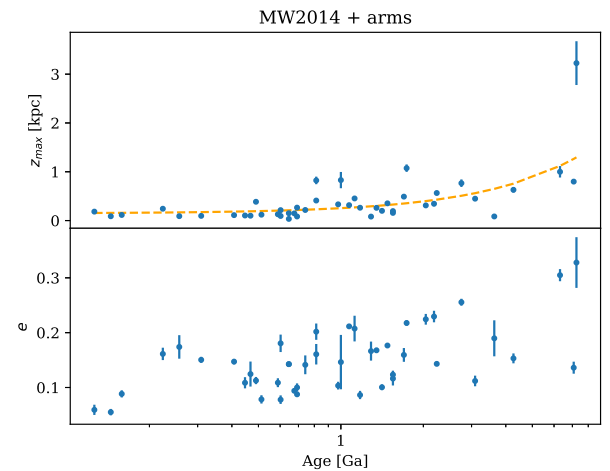}
\caption{As Fig.~\ref{fig:zmax_e-age_bar0} but adding the potential of the spiral arms.}
\label{fig:zmax_e-age_bar2}
\end{figure}

In this appendix we present figures similar to Fig.~\ref{fig:zmax_e-age} but only with the axisymmetric potential (Fig.~\ref{fig:zmax_e-age_bar0}), the axisymmetric potential and bar (Fig.~\ref{fig:zmax_e-age_bar1}), and the axisymmetric potential and spiral arms (Fig.~\ref{fig:zmax_e-age_bar2}), respectively.

\section{observed stars}\label{apex:observed-stars}

The complete list of observed stars will be public in CDS.

\begin{table}
	\label{tab:indiv_stars}
\end{table}

\clearpage

\end{document}